\pdfoutput=1
\documentclass[cernpreprint, atlasdraft=false, UKenglish, texmf, orcidlogo]{atlasdoc}
\usepackage{atlaspackage}
\usepackage{atlasbiblatex}

\usepackage{atlasphysics}
\usepackage{adjustbox}
\usepackage{multirow}
\usepackage{float}
\floatstyle{plaintop}
\restylefloat{table}
\floatstyle{plain}
\restylefloat{figure}

\graphicspath{{logos/}{figures/}}

\usepackage{ANA-HMBS-2024-64-PAPER-defs}

\AtlasTitle{Searches for direct slepton production in the compressed-mass corridor in $\sqrt{s}=13$ TeV $pp$ collisions with the ATLAS detector}

\AtlasAbstract{%
This paper presents searches for the direct pair production of charged light-flavour sleptons, each decaying into a stable neutralino and an associated Standard Model lepton.
The analyses focus on the challenging ``corridor'' region, where the mass difference, $\Delta m$, between the slepton ($\tilde{e}$ or $\tilde{\mu}$) and the lightest neutralino ($\tilde{\chi}^{0}_{1}$) is less or similar to the mass of the $W$ boson, $m(W)$, with the aim to close a persistent gap in sensitivity to models with $\Delta m \lesssim m(W)$.
Events are required to contain a high-energy jet, significant missing transverse momentum, and two same-flavour opposite-sign leptons ($e$ or $\mu$).
The analysis uses $pp$ collision data at $\sqrt{s} = 13$~TeV recorded by the ATLAS detector, corresponding to an integrated luminosity of 140~fb$^{-1}$.
Several kinematic selections are applied, including a set of boosted decision trees. These are each optimised for different $\Delta m$ to provide expected sensitivity for the first time across the full $\Delta m$ corridor.
The results are generally consistent with the Standard Model, with the most significant deviations observed with a local significance of $2.0\,\sigma$ in the selectron search, and $2.4\,\sigma$ in the smuon search.
While these deviations weaken the observed exclusion reach in some parts of the signal parameter space, the previously present sensitivity gap to this corridor is largely reduced.
Constraints at the 95\% confidence level are set on simplified models of selectron and smuon pair production, where selectrons (smuons) with masses up to 300 (350) GeV can be excluded for $\Delta m$ between 2 GeV and 100 GeV.

}

\AtlasRefCode{ANA-HMBS-2024-64}

\PreprintIdNumber{CERN-EP-2025-039}

\AtlasJournal{JHEP}
\AtlasJournalRef{JHEP 08 (2025) 053}
\AtlasDOI{10.1007/JHEP08(2025)053}

\hypersetup{pdftitle={ATLAS document},pdfauthor={The ATLAS Collaboration}}

\begin{document}

\maketitle

\tableofcontents

\clearpage

\section{Introduction}
%

% The next lines are included from the .//sleptons/introduction.tex input file
\FloatBarrier
The Large Hadron Collider (LHC) collaborations have performed a wide variety of searches for various new particles and phenomena.
Nevertheless, there remain models in which sensitivity to particles,
even those with electroweak-scale mass and coupling, is limited, in some cases barely exceeding that which was achieved at the Large Electron-Positron (LEP) collider.

Examples of these difficult new-physics scenarios are found in supersymmetric models. Supersymmetry (SUSY) is
an extension of the Standard Model (SM) that
posits an additional broken symmetry between fermions and bosons. It predicts additional
particles that share the same quantum numbers as their SM partners but differ by a half
unit of spin~\cite{myGolfand:1971iw, myVolkov:1973ix, Wess:1974tw, Wess:1974jb, Ferrara:1974pu, Salam:1974ig}.
The electroweak sector of SUSY consists of the superpartners of the SM leptons and the electroweak gauge and Higgs bosons.
The superpartners of the charged SM lepton fields are the selectron (\selectron), smuon (\smuon), and stau (\stau),
collectively referred to as ``sleptons''.
In this \paper{}, the sleptons of interest, denoted by \slepton, are restricted to refer
only to the first two generations: $\slepton\in\{\selectron,\,\smuon\}$.
In contrast to their fermionic superpartners, the sleptons are scalar particles.

The superpartners of the electroweak gauge fields are fermionic fields denoted bino ($\widetilde{B}$)
and wino ($\widetilde{W}$), which mix with the fermionic superpartners of the two SM Higgs doublets
($\widetilde{H}_{u,d}$) to form neutral and charged mass eigenstates.
In the Minimal Supersymmetric Standard Model (MSSM)~\cite{Fayet:1976et, Fayet:1977yc, myDjouadi:1998di, myBerger:2008cq},
after mixing, the resulting four neutral particles are called neutralinos and are denoted by $\tilde{\chi}^0_{1,2,3,4}$,
with the subscripts indicating increasing mass.

\begin{figure}[tp]
\centering
\includegraphics[width=0.35\textwidth]{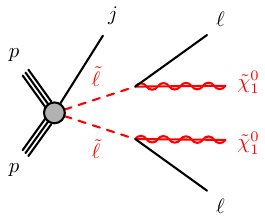}
\caption{
Diagram representing the direct production of sleptons (\slep) decaying into their
corresponding SM lepton partner ($\ell$) and the lightest neutralino (\None),
in association with a jet ($j$) from initial-state radiation.
}
\label{fig:SignalModel:feynmanDiagram}
\end{figure}

While searches for supersymmetric particles have not yet yielded discoveries,
SUSY remains an attractive framework for physics beyond the SM (BSM) since it can provide gauge coupling
unification~\cite{Sakai:1981gr, Dimopoulos:1981yj, Ibanez:1981yh, Dimopoulos:1981zb},
provide a mechanism for stabilising the Higgs boson mass at the weak scale~\cite{Barbieri:1987fn, mydeCarlos:1993yy},
resolve discrepancies between the measured and predicted values of the anomalous magnetic moment of the
muon~\cite{Bennett:2006fi, Hagiwara:2011af, Endo:2013bba, Ajaib:2015yma, PhysRevLett.131.161802},
and provide a viable dark matter candidate~\cite{SUSYDM}.

If sleptons and the \None{} are light in comparison to other SUSY particles,
then the dominant production mode at the LHC, and the dominant decay process, are the pair production of sleptons with $\slepton \to \ell \None$ decays, as shown in
Figure~\ref{fig:SignalModel:feynmanDiagram}.
Here, and later, $R$-parity~\cite{Farrar:1978xj} is assumed to be conserved, such that the \None is a stable particle that does not interact with the detector material. Therefore, its production can only be inferred from the presence of missing transverse momentum.
Another consequence of $R$-parity conservation is that supersymmetric particles must be produced in pairs,
with each one decaying into a final set of particles that includes the lightest supersymmetric particle (LSP).

The final state of interest comprises a pair $\ell^+\ell^-$ of opposite-charged leptons of the same flavour,
with missing transverse momentum $\ptmiss$ from the two unobserved $\None$ particles.
This two-lepton + $\ptmiss$ signal may be difficult to isolate --  in part because it can be
similar to SM production of $W^+W^-$ with decays into leptons and neutrinos.
A particularly difficult scenario occurs when the mass difference between slepton and neutralino is less or similar to the mass of the $W$ boson $m(W)$, i.e. $\dm \lesssim m(W)$, since then the signal and the $W^+W^-$ background distributions tend to be  kinematically more similar.
Another experimentally challenging scenario is one where $\dm$ is of the order of a few~\GeV,
since then only a small amount of energy is available in the \slep decay to emit leptons, which then may be too soft to satisfy trigger and reconstruction thresholds.

Searches for direct slepton production were previously undertaken
using leptonic final states at LEP and at the LHC\@.
The reported bounds on the slepton masses depend on the assumptions made.
Often, a four-fold mass degeneracy of the slepton mass is assumed for interpretational simplicity such that
$m(\selectronL)=m(\selectronR)=m(\smuonL)=m(\smuonR)$, where the subscripts indicate the chirality of the corresponding SM lepton.
In this scenario, the reported sensitivity is increased compared with those in which some of the sleptons are significantly heavier than others.
Since the SUSY mass breaking mechanism is not known there is no strong reason for assuming these four
sleptons to all have the same masses.
If the assumption of four-fold mass degeneracy is relaxed, as is permitted in the general MSSM, then the lower limit on the mass of any particular slepton, $m(\slep)$, can be significantly reduced or may even vanish for certain values of  $m(\slep,\None)$, leaving open  corridors  without experimental constraints. When bounds are reported on individual slepton states, the exclusions for \slepL tend to be stronger than those for \slepR, since the latter lacks SU(2) couplings and therefore has a lower production cross-section for the same mass.

The LEP experiments collectively place constraints on the $\selectronR$ and $\smuonR$ masses, excluding
$m(\selectronR) < 99.9\,\GeV{}$ and $m(\smuonR) < 95\,\GeV$, provided that $\dm\gtrsim 2~\GeV$~\cite{LEP-slepton}.
Constraints are also made on the $\selectron$ mass in the most mass-degenerate cases.
The ALEPH experiment scanned MSSM parameter space and provided an absolute lower limit $m(\selectronR)> 73\,\GeV$~\cite{ALEPH:2002nwp}.
There is no corresponding absolute lower limit from LEP on $m(\smuonR)$ for $\dm<2~\GeV$.

At the LHC, ATLAS has performed searches for slepton production using
lepton triggers~\cite{SUSY-2013-11,SUSY-2018-32} that achieved
lower limits on $m(\slepton)$ of up to 700\,\GeV{} when assuming four-fold mass degeneracy, while this lower limit reduces to
550\,\GeV{} for $m(\smuonL)=m(\smuonR)$ or $m(\selectronL)=m(\selectronR)$ (two-fold degeneracy).
To gain sensitivity to the region with $\dm \approx m(W)$, a dedicated ATLAS search was performed~\cite{SUSY-2019-02}.
However, the limits from these searches degrade at lower $\dm$, vanishing completely when $\dm \lesssim m(W)$.
To gain sensitivity to this region with smaller $\dm$,
other searches were performed at ATLAS triggering on the $\ptmiss$ signature,
permitting to select lower-\pT{} leptons~\cite{SUSY-2018-16}.
Limits were placed on $m(\slepton)$ in the range of
$0.5\,\GeV \lesssim \dm \lesssim 30\,\GeV$, with a $\Delta m$ dependent bound on individual slepton masses
that ranged from just larger than the LEP limits up to about 240\,\GeV.  %

The CMS Collaboration has reported sensitivity within a similar simplified model, again involving the direct pair production of
sleptons with four-fold mass degeneracy assumed throughout.
For $\dm \gtrsim m(W)$, bounds were set for values of $m(\slepton)\lesssim700\,\GeV$~\cite{CMS-SUS-20-001}.
To explore the more difficult region with $\dm\lesssim m(W)$,
the CMS Collaboration recasted the soft dilepton compressed electroweakino search~\cite{CMS-SUS-18-004}
as a slepton search~\cite{CMS-SUS-21-008} and obtained limits to exclude $m(\slep) \lesssim 215~\GeV$
for mass differences $\dm\approx 5~\GeV$, with limits for larger mass differences reducing to that of LEP at $\dm\approx 20~\GeV$.

These previous ATLAS and CMS analyses have provided improved sensitivity to light sleptons.
Nevertheless, there remains a striking gap in sensitivity in the challenging ``corridor'' region where
$\dm$ lies in the approximate range of 20\,\GeV{} to 60\,\GeV{}.
This region has proved sufficiently difficult that, for a substantial range of \dm, the LHC experiments have not yet
gained sensitivity beyond LEP\@, motivating phenomenological studies~\cite{PhysRevD.109.075018}.
The purpose of this analysis is to target that region of \dm, not necessarily assuming that the masses of all four
sleptons are equal.
Limits are set both on four-fold mass degenerate direct slepton production, and on the individual two-fold mass degenerate cases.

%
%
%
%
%
%

% End of text imported from the .//sleptons/introduction.tex input file

%
\section{ATLAS detector}
%

% The next lines are included from the .//sleptons/atlas-detector.tex input file
The ATLAS detector~\cite{PERF-2007-01} at the LHC covers nearly the entire solid angle around the collision point.\footnote{\AtlasCoordFootnote}
It consists of an inner tracking detector surrounded by a thin superconducting solenoid, electromagnetic and hadronic calorimeters,
and a muon spectrometer incorporating three large superconducting air-core toroidal magnets.

The inner-detector system (ID) is immersed in a \qty{2}{\tesla} axial magnetic field
and provides charged-particle tracking in the range \(|\eta| < 2.5\).
The high-granularity silicon pixel detector covers the vertex region and typically provides four measurements per track,
the first hit generally being in the insertable B-layer (IBL) installed before Run~2~\cite{ATLAS-TDR-19,PIX-2018-001}.
It is followed by the SemiConductor Tracker, which usually provides eight measurements per track.
These silicon detectors are complemented by the transition radiation tracker (TRT),
which enables radially extended track reconstruction up to \(|\eta| = 2.0\).
The TRT also provides electron identification information
based on the fraction of hits (typically 30 in total) above a higher energy-deposit threshold corresponding to transition radiation.

The calorimeter system covers the pseudorapidity range \(|\eta| < 4.9\).
Within the region \(|\eta|< 3.2\), electromagnetic calorimetry is provided by barrel and
endcap high-granularity lead/liquid-argon (LAr) calorimeters,
with an additional thin LAr presampler covering \(|\eta| < 1.8\)
to correct for energy loss in material upstream of the calorimeters.
Hadronic calorimetry is provided by the steel/scintillator-tile calorimeter,
segmented into three barrel structures within \(|\eta| < 1.7\), and two copper/LAr hadronic endcap calorimeters.
The solid angle coverage is completed with forward copper/LAr and tungsten/LAr calorimeter modules
optimised for electromagnetic and hadronic energy measurements respectively.

The muon spectrometer comprises separate trigger and
high-precision tracking chambers measuring the deflection of muons in a magnetic field generated by the superconducting air-core toroidal magnets.
The field integral of the toroids ranges between \num{2.0} and \qty{6.0}{\tesla\metre}
across most of the detector.
Three layers of precision chambers, each consisting of layers of monitored drift tubes, cover the region \(|\eta| < 2.7\),
complemented by cathode-strip chambers in the forward region, where the background is highest.
The muon trigger system covers the range \(|\eta| < 2.4\) with resistive-plate chambers in the barrel, and thin-gap chambers in the endcap regions.

The luminosity is measured mainly by the LUCID--2~\cite{LUCID2} detector that records Cherenkov light produced in the quartz windows of photomultipliers located close to the beampipe.

Events are selected by the first-level trigger system implemented in custom hardware,
followed by selections made by algorithms implemented in software in the high-level trigger~\cite{TRIG-2016-01}.
The first-level trigger accepts events from the \qty{40}{\MHz} bunch crossings at a rate close to \qty{100}{\kHz},
which the high-level trigger further reduces in order to record complete events to disk at about \qty{1.25}{\kHz}.

A software suite~\cite{mySOFT-2022-02} is used in data simulation, in the reconstruction
and analysis of real and simulated data, in detector operations, and in the trigger and data acquisition
systems of the experiment.

% End of text imported from the .//sleptons/atlas-detector.tex input file

\section{Analysis strategy}

% The next lines are included from the .//sleptons/strategy.tex input file
The searches reported in this \paper{} select events with a pair of same-flavour
electrons or muons with opposite electric charge.
They require the presence of a high-\pT{} jet from initial-state radiation (ISR) to
boost the neutralinos in the transverse plane.
This results in enough missing transverse momentum to trigger the events,
removing the need to trigger on leptons, hence extending sensitivity to smaller $\dm$.
Such ISR configurations also tend to boost the parent particles,
which helps distinguish the kinematics from those of the $W^+W^-$ background~\cite{Barr:2007hy},
and can increase the acceptance and efficiency for leptons from slepton decays with $\dm\lesssim m(W)$.

In addition to this selection, two complementary approaches are followed.
In the first approach, two signal regions are defined using a traditional `cut-and-count' strategy involving an optimised selection based on a subset of the kinematic variables
defined in Section~\ref{sec:SR}.
This approach is used to derive `model-independent' limits on event yields for BSM physics in the signal regions.

The second approach employs a set of five different Boosted Decision Trees (BDTs).
These BDTs are trained on slepton signal samples and on the SM backgrounds, with each BDT targeting signals within a relatively narrow range of $\dm$, and exploit a larger number
of kinematic variables compared with the cut-and-count-based analysis.
In combination, they account for differences in the signal kinematics that are driven by $\dm$, thereby providing wide-ranging discrimination between the signal and the backgrounds across the gap region.
Compared with the cut-and-count-based analysis, the BDTs exploit correlations between many kinematic variables and ultimately provide better expected sensitivity for the simplified signal model.

For both approaches, the signal regions are optimised based on the expected sensitivity for the simplified
supersymmetric models. In both cases, sensitivity to supersymmetric models in the gap region is found.
Exclusion limits are set using both approaches for the four-fold mass-degenerate slepton interpretation, while for the two-fold mass-degenerate sleptons, limits are determined using only the BDT approach.

% End of text imported from the .//sleptons/strategy.tex input file

%
\section{Data and simulated event samples}
%

% The next lines are included from the .//sleptons/samples.tex input file
%
This search uses $\sqrt{s} = 13~\TeV$ $pp$ collision data collected by the ATLAS experiment from 2015 to 2018,
where the minimum separation between consecutive bunch crossings was $\SI{25}{\ns}$ and the average number of interactions
per bunch crossing was approximately 33.7.
The application of data quality requirements~\cite{DAPR-2018-01} results in a data sample corresponding to an
integrated luminosity of 140~\ifb with an uncertainty of 0.83\%~\cite{DAPR-2021-01}, obtained using the LUCID-2 detector
for the primary bunch-by-bunch luminosity measurement, complemented by measurements using the ID and calorimeters.
The data events used in this search were collected using missing transverse momentum triggers~\cite{TRIG-2019-01} with varying
thresholds depending on the data-taking periods.
The triggers are found to be $>95\%$ efficient in all data-taking periods for $\met > 200~\GeV$, where $\met$ denotes the
magnitude of \ptmiss.

Monte Carlo (MC) simulations are used to estimate the event yields and systematic uncertainties for both signal processes and SM
backgrounds featuring prompt-lepton production.
The effect of multiple interactions in the same and neighbouring bunch
crossings (\pileup) was modelled by overlaying the simulated hard-scattering event with
inelastic proton--proton (\(pp\)) events generated with \PYTHIA[8.186]~\cite{Sjostrand:2007gs}
using the \NNPDF[2.3lo] set of parton distribution functions (PDF)~\cite{Ball:2012cx} and the
A3 set of tuned parameters~\cite{ATL-PHYS-PUB-2016-017}.
All MC events were then re-weighted to match the pile-up distribution observed in the data.
To simulate the detector response, background and signal MC samples were processed through the ATLAS simulation
framework~\cite{SOFT-2010-01} in \textsc{Geant4}~\cite{Agostinelli:2002hh}; the backgrounds made use of the
full detector simulation, while the signal samples used the ATLAS fast simulation that parameterises the response
of the calorimeters~\cite{ATL-PHYS-PUB-2010-013}.

The SM backgrounds involving the production of prompt leptons are modelled with a variety of MC generators.
Table~\ref{tab:mc_backgr} provides a comprehensive overview of all SM background samples utilised in the analysis.
It details the generators, the PDF sets, the sets of tuned parameters for the parton shower (including underlying event and hadronisation settings),
and the perturbative order in the strong coupling constant, $\alphas$, used for cross-section calculations.
Additional information regarding ATLAS simulations of \ttbar, single-top ($Wt$, $t$-channel, $s$-channel), multiboson, and boson-plus-jet processes is available in the
corresponding public notes ~\cite{ATL-PHYS-PUB-2018-009,ATL-PHYS-PUB-2016-004,ATL-PHYS-PUB-2017-005,ATL-PHYS-PUB-2017-006}.
The decays of bottom and charm hadrons are performed by \EVTGEN versions 1.2.0 and 1.6.0~\cite{Lange:2001uf}, except for the
backgrounds modelled using \SHERPA, for which the decays are performed internally.

% The next lines are included from the .//tables/BkgSamples.tex input file
%
\begin{table}[tbp]
\tiny
\centering
\caption{\label{tab:mc_backgr}Simulated background event samples with the corresponding matrix element and parton shower (PS) generators, cross-section order in $\alphas$ used to normalise the event yield, underlying-event tune and the generator PDF sets used. For diboson, triboson and $t\bar{t}{+}V$ samples, $V\in\{W,Z\}$. Diboson samples also include Higgs boson contributions. `Default' refers to the default tune of the \SHERPA generator.
Abbreviations used are defined as: leading-order (LO), next-to-leading-order (NLO), next-to-next-to-leading-order (NNLO), next-to-leading-logarithmic (NLL), next-to-next-to-leading-logarithmic (NNLL).}
\adjustbox{max width=\textwidth}{
\begin{tabular}{lllllll}
\toprule
\toprule

Physics process          			& Generator 	& Parton shower 	&  Normalisation 		& Tune 			& PDF (generator) & PDF (PS)\\
\midrule

$t\bar{t}$ 				& \POWHEGBOX[v2]~\cite{Frixione:2007nw,Nason:2004rx,Frixione:2007vw,Alioli:2010xd}       		& \PYTHIA[8.230]~\cite{Sjostrand:2014zea}  		& NNLO+NNLL~\cite{Beneke:2011mq,Cacciari:2011hy,Baernreuther:2012ws,Czakon:2012zr,Czakon:2012pz,Czakon:2013goa,Czakon:2011xx} 		& A14~\cite{ATL-PHYS-PUB-2014-021} & \NNPDF[3.0nlo]~\cite{Ball:2014uwa} & \NNPDF[2.3lo]~\cite{Ball:2012cx}\\
Single-top 				& \POWHEGBOX[v2]~\cite{Re:2010bp,Nason:2004rx,Frixione:2007vw,Alioli:2010xd} 	& \PYTHIA[8.230]	&NLO+NNLL~\cite{Kidonakis:2010ux,Kidonakis:2013zqa}	& A14 			& \NNPDF[3.0nlo]%
& \NNPDF[2.3lo]%
\\
Diboson $VV$			&  \SHERPA[2.2.2],\,2.2.11,\,2.2.12~\cite{Bothmann:2019yzt}	& 	\SHERPA[2.2.2],\,2.2.11,\,2.2.12~\cite{Gleisberg:2008fv,Schumann:2007mg}
& LO--NLO~\cite{Hoeche:2011fd,Hoeche:2012yf,Catani:2001cc,Hoeche:2009rj}   		&  Default~\cite{ATL-PHYS-PUB-2017-005}
&\NNPDF[3.0nnlo]   & \NNPDF[3.0nnlo] \\
Triboson $VVV$				& \SHERPA[2.2.2]			&  \SHERPA[2.2.2]		& NLO			& Default 	& \NNPDF[3.0nnlo]%
& \NNPDF[3.0nnlo]%
\\
$t\bar{t}{+}V$ 			& \MGNLO[2.3.3]~\cite{Alwall:2014hca}			& \PYTHIA[8.210]~\cite{Sjostrand:2014zea}						& NLO~\cite{Alwall:2014hca} 			& A14 & \NNPDF[3.0nlo]%
& \NNPDF[2.3lo]%
\\
$t\bar{t}{+}\gamma$ 			& \MGNLO[2.3.3]			& \PYTHIA[8.210]						& NLO~\cite{Melnikov:2011ta} 			& A14 & \NNPDF[2.3lo]%
& \NNPDF[2.3lo]%
\\
$t\bar{t}{+}H$ 				& \POWHEGBOX[v2]~\cite{Frixione:2007nw,Nason:2004rx,Frixione:2007vw,Alioli:2010xd,Hartanto:2015uka}      		& \PYTHIA[8.230] 		& NLO~\cite{mydeFlorian:2016spz} & A14 & \NNPDF[3.0nlo] & \NNPDF[2.3lo]\\
$t\bar{t}{+}WW$				& \MGNLO[2.2.2]			& \PYTHIA[8.186]~\cite{Sjostrand:2007gs} 		& NLO~\cite{Alwall:2014hca}			& A14			& \NNPDF[2.3lo]%
&  \NNPDF[2.3lo]%
\\  %
$t\bar{t}{+}WZ$				& \MGNLO[2.3.3]			& \PYTHIA[8.212]~\cite{Sjostrand:2014zea}%
& NLO~\cite{Alwall:2014hca}			& A14			& \NNPDF[2.3lo]%
&  \NNPDF[2.3lo]%
\\  %
$tZ,tWZ,t\bar{t}t\bar{t},t\bar{t}t$		& \MGNLO[2.3.3]			& \PYTHIA[8.230]%
& NLO~\cite{Alwall:2014hca}      		& A14			& \NNPDF[3.0nlo]%
& \NNPDF[2.3lo]%
\\
$Z/\gamma^*(\rightarrow \ell\ell,\tau\tau )$+jets 	& \SHERPA[2.2.11],\,2.2.14~\cite{Bothmann:2019yzt}			& \SHERPA[2.2.11],\,2.2.14~\cite{Schumann:2007mg}		& NNLO~\cite{Anastasiou:2003ds}			& Default	& \NNPDF[3.0nnlo] & \NNPDF[3.0nnlo] \\

ggF $H\rightarrow\tau\tau, \mu\mu$           &\POWHEGBOX[v2]~\cite{Hamilton:2013fea,Hamilton:2015nsa,Alioli:2010xd,Nason:2004rx,Frixione:2007vw}                    & \PYTHIA[8.2]~\cite{Sjostrand:2014zea}                  &NNLO~\cite{mydeFlorian:2016spz}              &AZNLO~\cite{STDM-2012-23}      & \PDFforLHC[15nnlo]~\cite{Butterworth:2015oua}   & \PDFforLHC[15nnlo] \\

VBF $H\rightarrow\tau\tau$           &\POWHEGBOX[v2]~\cite{Nason:2009ai,Alioli:2010xd,Nason:2004rx,Frixione:2007vw}                   & \PYTHIA[8.2]~\cite{Sjostrand:2014zea}                  &NNLO~\cite{Ciccolini:2007jr,Ciccolini:2007ec,Bolzoni:2010xr}              &AZNLO      & \PDFforLHC[15nlo]~\cite{Butterworth:2015oua}   & \PDFforLHC[15nlo] \\

$ZH, WH$           &\POWHEGBOX[v2]~\cite{Nason:2009ai,Alioli:2010xd,Nason:2004rx,Frixione:2007vw}                  & \PYTHIA[8.2]~\cite{Sjostrand:2014zea}                  &NNLO~\cite{myCiccolini:2003jy,Brein:2003wg,Brein:2011vx,Altenkamp:2012sx,Denner:2014cla,Brein:2012ne,Harlander:2014wda}              &AZNLO      & \PDFforLHC[15nlo]   & \PDFforLHC[15nlo] \\

\bottomrule
\bottomrule
\end{tabular}
}
\end{table}

% End of text imported from the .//tables/BkgSamples.tex input file

%
Simulated signal samples consisting of direct selectron (\selectronLR) and smuon (\smuonLR) pair production are used to optimise the event
selection and interpret the results.
During event generation, simplified models are generated assuming that selectrons and smuons are mass-degenerate: $m(\selectronL) = m(\selectronR) = m(\smuonL) = m(\smuonR)$.
The sleptons are required to decay into their corresponding SM lepton partner and a purely bino-like neutralino LSP, \chioz, with a branching ratio of 100\%.
For interpretations where the selectron-smuon mass degeneracy is lifted, no dependence of the event kinematics on the slepton flavour and chirality is expected, and the cross-sections are rescaled accordingly.
The events are produced at LO in $\alpha_\text{s}$ using \madgraph~2.7.3~\cite{Alwall:2014hca}
with the \NNPDF[2.3lo] PDF set and include up to two additional partons in the matrix element.
\PYTHIA[8.244]~\cite{Sjostrand:2014zea} is used to model the slepton decays, parton shower, hadronisation, and underlying event with the
A14 set of tuned parameters~\cite{ATL-PHYS-PUB-2014-021}.
Matching between the matrix element and parton shower is performed following the CKKW-L scheme~\cite{Lonnblad:2011xx}
with the merging scale set to $m(\sleptonLR)/4$.
The decays of bottom and charm hadrons are performed by \EVTGEN[1.7.0].
The signal production cross-sections and uncertainties are calculated using \textsc{Resummino} 2.0.1 at NLO+NLL
precision~\cite{Beenakker:1999xh,Debove:2010kf,Fuks:2012qx,Fuks:2013vua,Fiaschi:2018hgm,myBozzi:2007qr,Fuks:2013lya,Fiaschi:2018xdm}.
As an example, the combined production cross-sections for mass-degenerate selectrons and smuons with a mass of $200~\GeV$ are
$\sigma(\sleptonL) = 43.9 \pm 0.8$~fb and $\sigma(\sleptonR) = 16.7 \pm 0.4$~fb.

% End of text imported from the .//sleptons/samples.tex input file

%
\section{Object reconstruction}
%

% The next lines are included from the .//sleptons/reconstruction.tex input file
Each event is required to have a primary vertex built from at least two associated tracks with $\pT > 0.5~\GeV$.
The primary vertex with the highest sum of squared transverse momenta $\sum \pt^2$ of associated tracks~\cite{ATL-PHYS-PUB-2015-026} is selected as the hard-scatter vertex of interest in each event. A set of basic quality criteria is applied to ensure a fully operating detector and to reject events with detector noise or non-collision backgrounds~\cite{DAPR-2012-01}.

As described below, leptons and jets are “preselected” using loose identification criteria, and must survive tighter “signal” identification requirements to be selected for the search regions.
Preselected objects are used in fake/non-prompt (FNP) lepton background estimates and in resolving ambiguities between detector signals associated with multiple lepton and jet candidates.

Hadronic jets are reconstructed using the anti-$k_t$ algorithm~\cite{Cacciari:2008gp} as implemented in \texttt{FastJet}~\cite{Fastjet} with a jet radius parameter of $R=0.4$.
The inputs of this algorithm are particle-flow objects~\cite{PERF-2015-09} that combine measurements from the ATLAS inner detector and calorimeters~\cite{PERF-2014-07}
to improve the jet energy resolution and increase the jet reconstruction efficiency, especially at low jet \pt.
The jet energy scale and resolution are calibrated using simulations, with in situ corrections obtained from data~\cite{JETM-2018-05}.
Preselected jets, used for removing overlaps between different types of physics objects, are required to satisfy $\pt > 20~\GeV$ and $|\eta|<4.5$.
Signal jets, which are a subset of preselected jets and are used for event selection and categorisation, must satisfy stricter requirements that suppress contributions from pile-up.
All signal jets are required to have $\pt>30~\GeV$. Central signal jets with $|\eta|<2.4$ must additionally satisfy the \wp{Tight} working point of the jet vertex tagger~\cite{PERF-2014-03} if they have $\pt<60~\GeV$.
Jets within $|\eta|<2.5$ that satisfy the 77\% efficiency working point of the DL1r algorithm~\cite{FTAG-2019-07} are considered to contain $b$-hadrons and are referred to as $b$-tagged jets.

Preselected electrons are reconstructed using ID tracks matched to energy clusters in the electromagnetic calorimeter. These satisfy $\pt > 4.5$~\GeV{} and $|\eta| < 2.47$ with a \wp{LooseAndBLayerLLH} identification~\cite{EGAM-2018-01}.
The longitudinal impact parameter $z_0$ of preselected electron tracks is required to satisfy $|z_0\sin\theta| < 0.5$~mm.
Signal electrons must also satisfy the \wp{Medium} likelihood-based identification criteria
and have a transverse impact parameter $d_0$ with uncertainty $\sigma(d_0)$ satisfying $\left|d_0/\sigma(d_0)\right| < 5$.
To further reject non-prompt leptons, a multivariate isolation discriminant, similar to the one described in Ref.~\cite{HIGG-2017-02}, is employed that takes energy deposits and charged-particle tracks in a cone around the electron candidate as input.
The isolation criterion yields an efficiency between 80\% and 95\% for prompt electrons with \pt of 10~\GeV and 100~\GeV, respectively.

Preselected muons are reconstructed by combining tracks from the ID and the muon spectrometer subsystems.
The \wp{Medium} identification criteria~\cite{MUON-2018-03} are applied.  Preselected muons are required to have $\pt > 3~\GeV$ and $|\eta| < 2.7$,  and satisfy $|z_0\sin\theta| < 0.5$~mm.
Signal muons must have impact parameter significance $\left|d_0/\sigma(d_0)\right| < 3$ and must satisfy the multivariate-based isolation requirement, yielding efficiencies between 86\% and 98\% for prompt muons with \pt of 10~\GeV and 100~\GeV, respectively.

The missing transverse momentum \met is calculated as the magnitude of the negative vector sum of the transverse momenta of all identified hard physics objects (preselected leptons and jets) calibrated to their respective energy scales, with a contribution from an additional soft term~\cite{JETM-2020-03}.
This soft term is constructed from ID tracks matched to the hard-scatter vertex but not associated with any of the hard reconstructed objects. An object-based missing transverse momentum significance is used to identify events in which the reconstructed \met comes from undetected particles rather than mismeasured energy deposits.

To prevent the use of the same reconstructed detector signals in multiple objects, an overlap removal procedure is applied to the preselected leptons and jets in the following order.
First, any electron sharing an ID track with a muon is removed. Next, jets are removed if they are within $\Delta R < 0.2$ from a remaining electron.
After this, electrons are in turn rejected if they are within $ 0.2 < \Delta R < \min\left(0.4, 0.04 + 10~\GeV / \pt(e)\right)$ of any remaining jet.
Jets are removed if they are closer than $\Delta R < 0.2$ to a muon and the jet has fewer than three associated tracks with $\pT > 500 ~\MeV$.
Finally, any muon within $0.2 < \Delta R < \min\left(0.4, 0.04 + 10~\GeV / \pt(\mu)\right)$ of a jet is removed.

% End of text imported from the .//sleptons/reconstruction.tex input file

%
\section{Event selection}
\label{sec:SR}

% The next lines are included from the .//sleptons/selection.tex input file
Within the simplified model, the produced sleptons are assumed to decay according to the process $\tilde{\ell} \rightarrow \tilde{\chi}^0_1 + \ell$ with a $100\%$ branching ratio.
The final-state signature of this process is two same-flavour, opposite-sign (SFOS) leptons, and missing transverse momentum from the neutralinos.
The event selection is designed to take advantage of this decay signature, and the additional requirement of an ISR jet that boosts the decay products, to increase the \met.
For the cut-and-count-based approach, the event selection is performed by imposing requirements on kinematic variables with good separation between the signal and the SM backgrounds.
In the BDT approach, the event selection is carried out using BDTs trained to distinguish the signal from the SM background using a set of discriminating kinematic variables as inputs.
The variables used by the cut-and-count-based and BDT approaches are defined below.
The leading lepton (jet) is defined as the lepton (jet) with the largest \pt in the event. The superscript \markvar indicates that a variable was used in the BDT training (see Section~\ref{sec:BDTSR}).

\begin{itemize}
\item The leading (\ptlepone\markvar) and subleading (\ptleptwo\markvar) lepton transverse momenta.
\item The invariant mass of the two leptons, \mll\markvar.
\item The angular distance between the two leptons, \dRll\markvar.
\item The modulus of the vector sum of the $\ptvec$ of the two leptons, \ptll\markvar.
\item The $\pt$ of the leading ($\pt^{\jetone}\markvar$) and subleading ($\pt^{\jettwo}\markvar$) jet.
\item The number of signal jets with $\pt > 30~\GeV$, $\njet^{30}$.
\item The number of \btagged jets with $\pt>20~\GeV$, $\nbjet^{20}$.
\item The magnitude of the missing transverse momentum vector, \met\markvar.
\item The azimuthal angle between the leading jet and the \ptmiss, \dPhiJetOneMET\markvar.
\item The minimum azimuthal angle between any signal jet and the \ptmiss, \mindPhiJetsMET.
\item The azimuthal angle between \ptmiss and the leading \dPhiLepOneMET\markvar and subleading lepton \dPhiLepTwoMET\markvar.
\item The object-based significance of the missing transverse momentum, \metsig\markvar, which provides a measure of the likelihood that the reconstructed \met originated from real invisible particles and not from detector effects.
\item The angular variable $\cos\theta^{*}_{\ell\ell}\markvar$, which is a measure of the polar angle of one of the two leptons with the beam axis in the laboratory frame and is sensitive to the spin of the particles produced in the collision~\cite{Barr:2005}.
Similarly, $\cos\theta^{V}_{\ell\ell}\markvar$ measures the polar angle of one of the two leptons with the beam axis, evaluated in the dilepton centre of mass frame~\cite{Melia_2012}.
As sleptons are scalar particles, these variables exploit shape differences relative to the SM backgrounds.
\item An approximation of the invariant mass of a pair of leptonically decaying $\tau$-leptons, denoted $M_{\tau\tau}$\markvar, assuming that the neutrinos are collinear with their parent $\tau$-lepton~\cite{Barr_2015,Baer_2014}.
This variable utilises the lepton \pt and the \MET to discriminate between backgrounds containing boosted $Z\to\tau\tau$ decays, for which $M_{\tau\tau}$ peaks near the $Z$ boson mass, and the signal processes that do not feature such a peak.
\item The transverse mass of the leading (\mtlepone\markvar) and subleading (\mtleptwo\markvar) lepton, defined as:
\begin{equation}
m_\text{T}(\pT,q_\text{T},m_{\chi}) = \sqrt{m_{\ell}^{2} + m^{2}_{\chi} + 2 \left( E^{\ell}_\mathrm{T} E^{q}_\mathrm{T} - \mathbf{p}_\mathrm{T} \mathbf{q}_\mathrm{T} \right) },
\end{equation}
where $m_\ell$ is the mass of the lepton, $m_{\chi}$ is the invisible particle mass (taken to be zero for the calculation of \mtlepone and \mtleptwo), and $E_\text{T}^{\ell}$ and $E_\text{T}^{q}$ are defined as: $E_\text{T}=\sqrt{m^2+ \mathbf{p}_\text{T}^2}$, for the lepton and invisible particle respectively.
The two vectors $\mathbf{p}_\text{T}$ and $\mathbf{q}_\text{T}$ are the transverse momentum vector of the lepton and the invisible particle \ptmiss vector, respectively.
\item The stransverse mass (\mttwo), a kinematic variable that is analogous to $m_\text{T}$, but in the case of a pair of particles that have both decayed into a visible and an invisible particle ~\cite{Lester:1999tx,Barr:2003rg}.
It is defined as:
\begin{equation*}
\mttwo^\chi
( \mathbf p_{\mathrm{T,1}}, \mathbf p_{\mathrm{T,2}},\mathbf p^\mathrm{miss}_\mathrm{T}) = \min_{\mathbf q_{\mathrm{T,1}} + \mathbf q_{\mathrm{T,2}} =
\mathbf p^\mathrm{miss}_\mathrm{T} }
\left\{ \max[\;
m_{\mathrm{T}}( \mathbf p_{\mathrm{T,1}}, \mathbf q_{\mathrm{T,1}};m_{\chi} ),
m_{\mathrm{T}}( \mathbf p_{\mathrm{T,2}}, \mathbf q_{\mathrm{T,2}};m_{\chi}
)
\;]
\right\},
\end{equation*}
where $m_{\mathrm T}$ indicates the transverse mass, as defined above.
The two vectors $\mathbf p_{\mathrm{T,1}}$ and $\mathbf p_{\mathrm{T,2}}$ are the transverse momentum vectors of the two leptons, and $\mathbf q_{\mathrm{T,1}}$, $\mathbf q_{\mathrm{T,2}}$ are vectors such that
$\mathbf p^\mathrm{miss}_\mathrm{T}= \mathbf q_{\mathrm{T,1}} + \mathbf q_{\mathrm{T,2}}$.
The mass of the invisible particle, $m_{\chi}$, is a free parameter of the equation.
A $\chi$ mass of zero corresponds to the SM neutrino and this variant is useful to reject backgrounds from $W^+W^-$ and \ttbar decays.
Non-zero values of $m_{\chi}$ are also chosen to reflect the signal model(s) with that invisible particle mass.
A key feature of this variable is a kinematic endpoint, dependent on the true invisible particle mass of the event.
Since the signal and SM background have different kinematic endpoints, this feature can be exploited in defining the cut-and-count-based signal regions.
In total, six different values of $m_{\chi}$ in the range 0 to 300~\GeV are employed to construct $\mttwo^{0}\markvar,\mttwo^{50\:}\markvar,\mttwo^{100\:}\markvar,\mttwo^{150\:}\markvar,\mttwo^{200\:}\markvar$ and $\mttwo^{300\:}$\markvar.
\end{itemize}

\subsection{Preselection requirements}

The cut-and-count and BDT-based approaches share several common preselection requirements, across all regions of the analysis, as summarised in Table~\ref{tab:CommonCuts}.
Events are required to have $\met > 200~\GeV$ to be in the plateau region of the used triggers and effectively reject SM backgrounds with no sources for real \met such as $Z\to\ell\ell \ +$ jets processes, where $\ell \in \{e, \mu\}$.
Events with additional leptons are vetoed to reduce contamination from multiboson backgrounds, in particular $WZ$. Multijet backgrounds in which the \met originates mainly from mismeasured jets are suppressed by requiring $\mindPhiJetsMET > 0.4$.
Requirements on $\pt^{\jetone}$ and \dPhiJetOneMET are used to select events with an ISR-like topology as present in the signal scenarios of interest.
As the leptons from the targeted signal originate from two different decay legs they are expected to be reasonably well separated from each other motivating a lower requirement on $\Delta R_{\ell \ell}$.
This selection is tightened further, individually for the cut-and-count and BDT-based channels.
These additional selection requirements are stricter in the cut-and-count approach than in the multivariate BDT approach where more relaxed criteria allow the BDTs to better exploit correlations between the input variables.
The signal regions (SRs) of the cut-and-count and BDT-based stategies are defined by the additional requirements described in Sections \ref{sec:CCSR} and \ref{sec:BDTSR}, respectively.

% The next lines are included from the .//tables/CommonPreselection.tex input file
\begin{table}[tbp]
\centering
\renewcommand{\arraystretch}{1.1}
\caption{Summary of the requirements applied to all cut-and-count and BDT-based regions of the analysis.}
\label{tab:CommonCuts}
\begin{tabular}[h]{cc}
\toprule
\toprule
\multicolumn{2}{c}{Common preselection requirements} \\
\midrule
$\met$ & $>200~\GeV$ \\
\nlepbase & $=2$ \\
\nlepsig & $=2$ \\
\mindPhiJetsMET & $> 0.4$ \\
\dPhiJetOneMET & $> 2.0 $ \\
$\dRll $ & $>0.75$ \\
$\njet^{30}$ & $>0$\\
$\pt^{\jetone} $ & $>100~\GeV $ \\
\midrule
\multicolumn{2}{c}{Additional requirements for cut-and-count-based analysis} \\
\midrule
$\pt^{\ell_{1}} \ \text{and} \ \pt^{\ell_{2}}$ & $>10~\GeV$  \\
$\njet^{30}$ & $<3$\\
\midrule
\multicolumn{2}{c}{Additional requirements for BDT-based analysis} \\
\midrule
$\pt^{\ell_{1}} \ \text{and} \ \pt^{\ell_{2}}$ & $>6~\GeV$  \\
\bottomrule
\bottomrule

\end{tabular}
\end{table}

% End of text imported from the .//tables/CommonPreselection.tex input file

\FloatBarrier

\subsection{Cut-and-count event selection}
\label{sec:CCSR}
In addition to the common preselection requirements, the cut-and-count-based approach implements additional tighter selections.
These are chosen to reduce backgrounds with a similar signature to the signal models and were optimised using distributions of sensitive kinematic variables and evaluating the expected significance.
The optimisation targeted in particular signal scenarios with mass splittings \dm between 20~\GeV and 40~\GeV, where events with low \pt leptons were found to provide little sensitivity due to the large FNP lepton background.
Similarly, events with three or more jets are substantially contaminated by the top background. Hence, such events are rejected in all cut-and-count based selections as reported in Table~\ref{tab:CommonCuts}.
Furthermore, events are required to contain a pair of SFOS leptons, $e^{\pm}e^{\mp}$ or $\mu^{\pm}\mu^{\mp}$, as present in the targeted signal models.
The optimisation of the event selection led to two sets of SRs, defined in Table \ref{tab:CCSRCuts}.
While both SRs share the same requirements on \met, $\nbjet^{20}$ \mtleptwo and a $Z$ boson veto based on \mll, \SRhigh (\SRlow) uniquely requires $\mtlepone > 100\,\GeV$ ($\mtlepone < 100\,\GeV$) to specifically target signals with large (small) values of \dm.
These SRs are further split by a di-electron or di-muon lepton-flavour requirement, which yields four SRs in total: \SRlowe, \SRlowm, \SRhighe, \SRhighm.
The targeted signal models have a kinematic endpoint in the $\mttwo$ distribution dependent on $\Delta m$, and this feature can be exploited by binning the SRs in this variable.
Thus, each of the four SRs are split into eight exclusive $\mttwo^{100}$ bins, with the following bin edges: $[100, 105, 110, 120, 130, 140, 150, 160, 180]~\GeV$.
The bin edges are more closely spaced at lower values of $\mttwo^{100}$ to increase sensitivity to signals with smaller mass-splittings.

% The next lines are included from the .//tables/CCSRCuts.tex input file
\begin{table}[tbp]
\centering
\caption{Definitions of the cut-and-count based SRs. The common preselection cuts in Table~\ref{tab:CommonCuts} are applied as well.}

\renewcommand{\arraystretch}{1.1}
\begin{tabular}[h]{ccc}
\toprule
\toprule
Variable & \SRhigh & \SRlow\\
\midrule
Lepton charge and flavour                         & \multicolumn{2}{c}{SFOS} \\
$\nbjet^{20}$                  & \multicolumn{2}{c}{$=0$} \\
$\met$                         & \multicolumn{2}{c}{$>300 \ \GeV$}\\
$|\mll - m_{Z}|$         & \multicolumn{2}{c}{ $>10 \ \GeV$} \\
$\mtlepone$              & $>100 \ \GeV$      & $<100 \ \GeV$ \\
$\mtleptwo$              & \multicolumn{2}{c}{$>100 \ \GeV$}  \\
\bottomrule
\bottomrule
\end{tabular}
\label{tab:CCSRCuts}
\end{table}

% End of text imported from the .//tables/CCSRCuts.tex input file

In addition to the exclusive SRs bins introduced above, the phase space of \SRhigh is further used to define a set of eight ``inclusive'' single-bin cut-and-count SRs referred to as \SRDisc. This set of SRs is used to probe the $\mttwo^{100}$ spectrum for an excess in a more model-independent way that does not require an assumption on the signal shape.
These SRs are defined to be increasingly inclusive in $\mttwo^{100}$ using the same bin edges as \SRhigh. For example, $\SRDisc^{< 105}$ covers $\mttwo^{100} \in [100, 105]~\GeV$ and $\SRDisc^{< 180}$ covers $\mttwo^{100} \in [100, 180]~\GeV$, i.e. $\SRDisc^{< 180}$ completely contains $\SRDisc^{< 105}$.
The \SRDisc are not split into lepton flavour, i.e. both $ee$ and $\mu\mu$ events are considered.

\subsection{BDT event selection}
\label{sec:BDTSR}

In this approach, event selections are based on the output from BDTs that are trained to distinguish signal from background events.
The training sample is based on MC events satisfying the common preselection requirements summarised in Table~\ref{tab:CommonCuts}.
Events containing leptons with $\pt < 6~\GeV$ are found to contribute little to the sensitivity for the signal of interest due to the large FNP lepton background and are consequently rejected in all BDT-based selections.
While lower-\pt leptons can in principle improve sensitivity to scenarios with mass splittings below 5~\GeV, this analysis is not optimised for such regimes which lie beyond its primary target of $\dm \gtrsim 10~\GeV$.
Further requirements are employed such that the BDT optimisation does not focus on events that are simple to separate.
Hence, training is restricted to SFOS events with no $b$-tagged jets and that satisfy a veto of on-shell $Z$ boson decays via $|\mll - m_{Z}| >10~\GeV$.
The training selection is inclusive of lepton flavour, as the di-electron and di-muon selections are similar kinematically.
Specifically, the BDT training requirements are a combination of the common requirements in Table \ref{tab:CommonCuts}, and the general requirements in Table \ref{tab:SRCuts}.
The normalised, pre-fit SM distributions of two kinematic variables, $\mttwo^{100}$ and \mtlepone, which show potential to discriminate between signal and SM backgrounds, are illustrated in Figure~\ref{fig:BDTPre}.
Overlaid are the kinematic distributions of three example SUSY models, corresponding to $m(\tilde{\ell}, \chioz) = (200, 180)~\GeV$, $(200, 170)~\GeV$, and $(200, 150)~\GeV$.

\begin{figure}[tbp]
\centering
\subfloat[]{
\includegraphics[width=0.45\columnwidth]{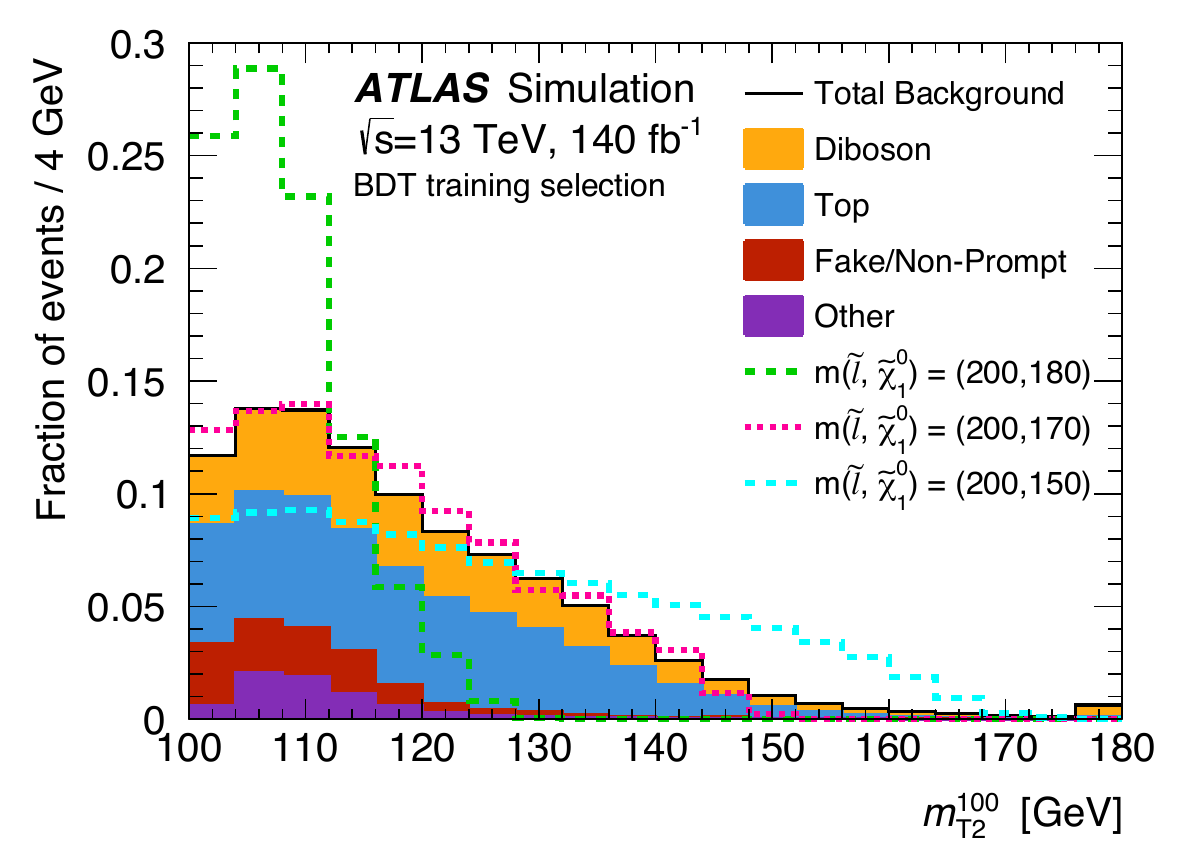}
}
\subfloat[]{
\includegraphics[width=0.45\columnwidth]{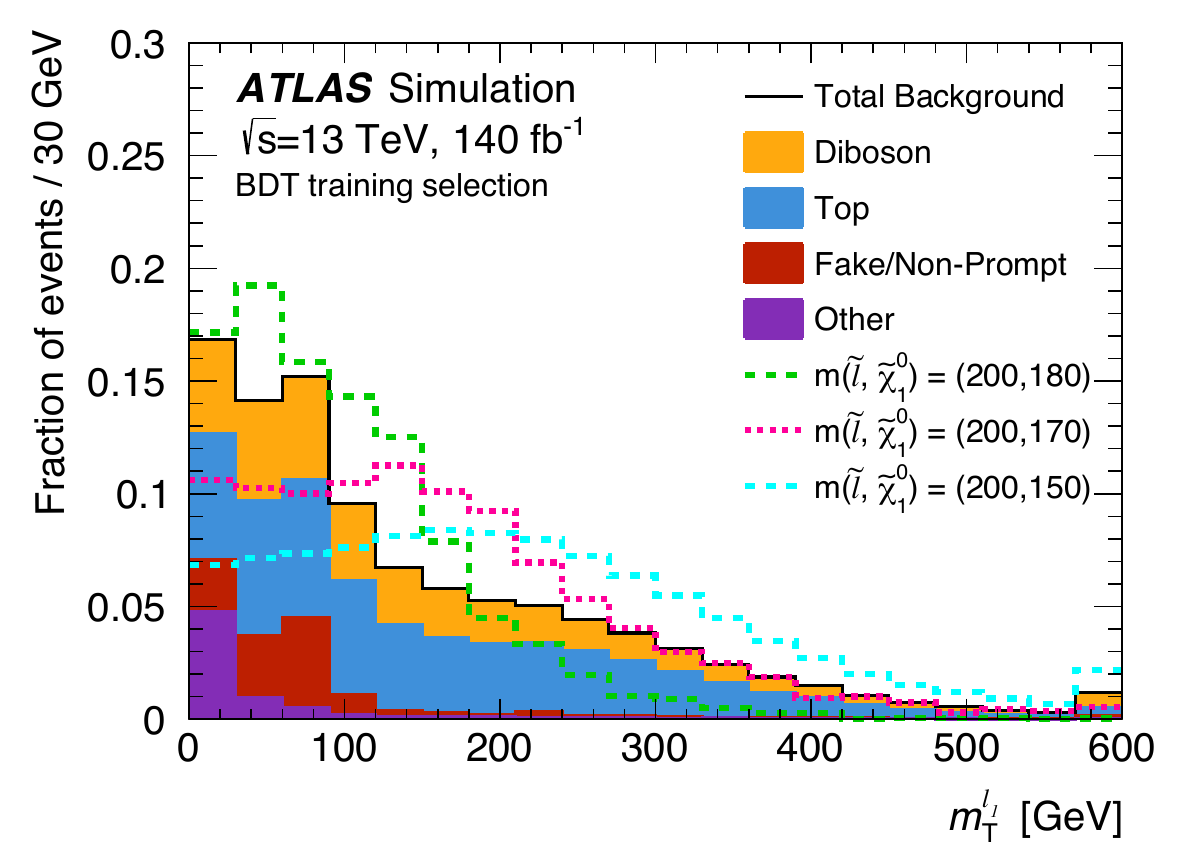}

}
\caption{Pre-fit SM background distributions of (a) $\mttwo^{100}$ and (b) \mtlepone normalised to unity using the selection to train the BDTs as described in the main text.
Three example SUSY scenarios are overlaid for illustration that correspond to $m(\tilde{\ell}, \chioz) = (200, 180)~\GeV$, $(200, 170)~\GeV$ and $(200, 150)~\GeV$. The last bin contains the overflow.}
\label{fig:BDTPre}
\end{figure}
Using these preselection requirements, five different BDTs are trained using the XGBoost package~\cite{XGBoost}, according to the mass-splitting of the targeted signal models.
The signal samples in the training step are grouped as follows, where the subscript denotes the grouping of the mass-splittings in that BDT: $\dmBase_{5+10}$, $\dmBase_{20}$, $\dmBase_{30}$,  $\dmBase_{40+50}$, $\dmBase_{60+75}$.
Signal kinematics change significantly with the mass-splitting, and use of multiple BDTs was found to give strong sensitivity across the entire range of targeted \dm.
Studies were carried out on grouping samples with neighbouring mass-splittings, and an optimal grouping of mass-splittings was determined through comparison of the classification strength of different groupings.
The groupings of the signal samples for the training of each BDT are shown in Appendix~\ref{sec:BDTTraining}.

All BDTs were trained using the same set of 23 kinematic variables described above that represent a mix of low- and high-level variables.
These were chosen by examining variables important to the cut-and-count-based approach and to previous analyses targeting similar final states~\cite{SUSY-2018-32,SUSY-2018-16,SUSY-2019-02}.
To prevent overtraining, early stopping is employed, where training is halted if performance on the validation set does not improve after 20 additional trees.
Hyperparameter optimisation was carried out for each $\dm$ BDT separately, to achieve optimum classification performance while preventing overtraining.
Details related to the hyperparameters and their final values can be found in Appendix \ref{sec:BDTTraining}.

A five-fold cross-validation strategy is employed, wherein the MC samples for each background process and signal group are divided into fifths.
Five BDTs are trained for each mass splitting, with each using three fifths of the data for training, one fifth for validation (to monitor overtraining), and one fifth for evaluation. This ensures no overlap between training and evaluation data across the five BDTs.
For each mass splitting, the final BDT score for each event is taken from the BDT for which the event was in the evaluation fifth, ensuring that each BDT is only used to evaluated events not used in its training.
The split is achieved by taking the modulus of the event number relative to five, and the same strategy is applied when determining which of the five BDTs used to evaluate the samples (e.g., data) not used in the training.

According to metrics provided by the SHAP package~\cite{SHAP}, the most important inputs used in the training were found to be the \metsig, the transverse mass calculated with the subleading lepton \mtleptwo and the \mttwo variables.
The importance ranking slightly varies across the BDTs and a more detailed overview can be found in Appendix~\ref{sec:BDTTraining}.

The BDT output, a classification score between 0 (background-like) and 1 (signal-like), is used to distinguish signal and background processes.
The definitions for the SRs of the BDT approach are shown in Table~\ref{tab:SRCuts}, and are applied in addition to the training selection described above. The tail of the BDT score is binned into three orthogonal SR bins to ensure sensitivity to the full range of signals, as the BDT score distributions peak at different values depending on the slepton mass. The binning was optimised individually for each set of BDTs to account for the different shape of the score distributions. Each SR bin is further split via the lepton flavour into $ee$ and $\mu\mu$ events yielding a total of six SRs for each BDT\@.
The SR naming convention follows the template: $\mathrm{SRX\mhyphen BDT}^{\mathrm{Z}}_{\mathrm{Y}}$, where X is the SR number $(1,2,3)$, Y is the BDT used to define the region, and Z is the lepton-flavour requirement.
For example, the first di-electron SR, defined using the smallest mass-splitting BDT is denoted by \VLowSROnee.

% The next lines are included from the .//tables/BDTSRDefinitions.tex input file
\begin{table}[tbp]
\centering
\renewcommand{\arraystretch}{1.1}
\caption[]{Definition of the BDT-based SR requirements that are applied in addition to the preselection requirements shown in Table~\ref{tab:CommonCuts}.
A BDT score requirement that is exclusive of the lower bound is indicated by the use of ``('' in defining the BDT score requirements.
Along with the BDT-score requirement, the SRs are further split into three opposite-sign di-electron SRs, and three opposite-sign di-muon SRs.}
\begin{tabular}[h]{cccc}
\toprule
\toprule
\multicolumn{4}{c}{General requirements} \\
\multicolumn{4}{c}{SFOS leptons \hfill $\nbjet^{20} = 0$ \hfill $|\mll - m_{Z}| >10~\GeV$} \\
\noalign{\vskip 2mm}
\midrule
\multicolumn{4}{c}{BDT Score requirements} \\
\multicolumn{1}{c|}{BDT set} &  \multicolumn{1}{c}{$\mathrm{SR1}^{\mathrm{ee/\mu\mu}}$}  & \multicolumn{1}{c}{$\mathrm{SR2}^{\mathrm{ee/\mu\mu}}$}  & \multicolumn{1}{c}{$\mathrm{SR3}^{\mathrm{ee/\mu\mu}}$}\\
\midrule
\multicolumn{1}{c|}{$\mathrm{BDT}_{{5+10}}$} & \multicolumn{1}{c}{$[0.93,0.953]$}  & \multicolumn{1}{c}{$(0.953,0.976]$}  & \multicolumn{1}{c}{$(0.976,1.0]$}\\
\multicolumn{1}{c|}{$\mathrm{BDT}_{{20}}$} & \multicolumn{1}{c}{$[0.90,0.928]$}  & \multicolumn{1}{c}{$(0.928,0.956]$}  & \multicolumn{1}{c}{$(0.956,1.0]$}\\
\multicolumn{1}{c|}{$\mathrm{BDT}_{{30}}$} & \multicolumn{1}{c}{$[0.88,0.915]$}  & \multicolumn{1}{c}{$(0.915,0.95]$}  & \multicolumn{1}{c}{$(0.95,1.0]$}\\
\multicolumn{1}{c|}{$\mathrm{BDT}_{{40+50}}$} & \multicolumn{1}{c}{$[0.90,0.928]$}  & \multicolumn{1}{c}{$(0.928,0.956]$}  & \multicolumn{1}{c}{$(0.956,1.0]$}\\
\multicolumn{1}{c|}{$\mathrm{BDT}_{{60+75}}$} & \multicolumn{1}{c}{$[0.90,0.935]$}  & \multicolumn{1}{c}{$(0.935,0.97]$}  & \multicolumn{1}{c}{$(0.97,1.0]$}\\

\bottomrule
\bottomrule

\end{tabular}
\label{tab:SRCuts}
\end{table}

% End of text imported from the .//tables/BDTSRDefinitions.tex input file

\FloatBarrier

% End of text imported from the .//sleptons/selection.tex input file

%
\FloatBarrier %
\section{Background estimation}
%

% The next lines are included from the .//sleptons/bkgEstimation.tex input file
The SM background can be categorised into irreducible and reducible backgrounds. Irreducible backgrounds originate from processes with prompt, real leptons that produce final states resembling the signal.
The dominant sources of irreducible background are processes containing top quark or diboson ($VV$) decays, the latter in particular from $W^+W^-$ events.
Events from the reducible background contain at least one non-prompt lepton from e.g. decays of hadrons containing heavy-flavour quarks or a ``fake'' lepton originating from misidentified detector signatures such as jets.
As lepton misidentification rates are typically small it would be computational expensive to model this FNP lepton with MC simulations with sufficiently large statistics. Hence, a data-driven method referred to as fake-factor method~\cite{EGAM-2019-01} is employed to estimate the FNP background.
In this analysis, the FNP background is dominated by $W\to\ell\nu + \text{jets}$ events, where one real lepton originates from the $W$ decay and the second lepton from a misidentified jet.

\subsection{Irreducible backgrounds}
In all signal selections, i.e. for the cut-and-count based and each of the BDT-based selections, two control regions (CRs) are used to constrain the two major sources of irreducible background. While a CR-VV is enriched in diboson processes, a  CR-Top is used to target processes involving top quarks.
As their composition was found to be similar between the CR and SR phase space, the \ttbar and single-top processes are estimated in conjunction via a common normalisation factor.
Subdominant irreducible backgrounds such as $Z\to\tau\tau+\text{jets}$ and multitop events were found to contribute typically less than 5\% in the SRs, and are taken directly from MC simulation. These are grouped into the ``other'' backgrounds category in the following, which consists of all non-$VV$, \ttbar and single-top background processes listed in Table~\ref{tab:mc_backgr}.

A dedicated set of CRs was constructed for the cut-and-count-based selection and for each of the BDT selections, which ensures that the SM backgrounds are constrained in a phase space close to but statistically independent of the respective SRs.
Consequently, background normalisations for the diboson and top sample, which get extracted by a simultaneous maximum-likelihood fit to the observed data in CR-VV and CR-top, are derived individually in each of the signal selections.
The derived background normalisations are then verified in a set of validation regions (VRs) that are also individually designed for the cut-and-count-based and BDT selections. In particular, changing the same-flavour to a different-flavour (DF) requirement on the leptons in the SRs results in orthogonal VR-DF regions that are otherwise identical to the SRs.
This takes advantage of the dominating, flavour symmetric $W^{+}W^{-}$  and \ttbar background processes producing SFOS and different-flavour, opposite-sign (DFOS) leptons at equal rates, whereas the signal model only produces SFOS leptons.

\subsection{Reducible backgrounds}

Reducible backgrounds attributed to FNP leptons can arise from several sources: jets misidentified as leptons, photon conversions into $e^+e^-$ pairs, and real, non-prompt leptons produced in the semileptonic decays of hadrons containing bottom or charm quarks.
From studies of simulated samples, the dominant sources of FNP electrons and muons in the SRs are light- and heavy flavour hadrons decays, respectively.

In the fake-factor method, the FNP contribution to a selection requiring tight, signal (`ID’) leptons such as the SRs is estimated from an orthogonal control sample failing to meet the tight criteria but satisfying looser (`anti-ID’) requirements on the leptons. The transfer factor to extrapolate from the loose to the tight FNP lepton sample is referred to as fake factor and depends on the lepton flavour and kinematics and hence a set of fake factors is derived to capture these dependencies. The fake factors are measured using data in a region that is enriched in FNP leptons and orthogonal to the regions in which they are applied.

In this analysis, ID leptons are identical to signal leptons, while anti-ID leptons are preselected leptons that fail to satisfy at least one of the criteria for signal leptons on identification, isolation and impact parameters. A control sample is constructed by applying the same kinematic criteria as the associated target region (which can by any CR, VR or SR) but requires one or both leptons to be of anti-ID quality. The contribution from irreducible backgrounds featuring two prompt leptons in the control sample is estimated from MC simulation and subtracted before the extrapolation.

The fake factors are measured in a data sample collected with prescaled low-\pt single-lepton triggers, which require looser selection criteria than those applied to anti-ID leptons.
This sample is referred to as the measurement sample and is dominated by multijet events with minor contributions from top and $W+\text{jets}$ processes.
The sample is required to contain one preselected lepton and is enriched in FNP leptons similar to those contaminating the target phase space by requiring the leading jet \pt to be greater than $100~\GeV$.
The fake factors are calculated as the ratio of the number of ID leptons to the number of anti-ID leptons in the measurement sample. To take into account the contamination from real leptons their contribution is estimated from \ttbar and $W+\text{jet}$ MC simulation and subtracted from the ID and anti-ID samples. The fake factors are measured individually for electrons and muons, and binned in lepton \pt to capture their dominant kinematic dependence.
Furthermore, there is a dependence of the fake factors on the absence or presence of a $b$-tagged jet in the event, which is attributed to the increased proportion of heavy flavour hadron decays in the latter.
Hence, the fake factors are further measured separately for events with and without a $b$-tagged jet.
The fake factors measured using events containing $b$-tagged jets are applied to the CRs and VRs that require $\nbjet^{20} > 0$ in order to constrain and validate the SM backgrounds involving top quark production, while the fake factors measured
in events without $b$-tagged jets are applied to all other analysis regions that require $\nbjet^{20} = 0$.
The measured fake factors in events without $b$-tagged jets range from approximately 0.1 (0.15) to 0.65 (0.1) for electrons (muon) with a \pt of around 10~\GeV and 50~\GeV, respectively.
The FNP estimates in the cut-and-count-based and each of the BDT selections are all based on the same set of fake factors.

To validate the data-driven FNP lepton estimate in the SRs, dedicated VRs are defined for both the BDT and cut-and-count-based approaches. These VRs impose a same-sign (SS) requirement on the charges of the lepton pair to be enriched in FNP background and are consequently referred to as VR-SS\@.
The kinematic requirements applied to each VR-SS are largely the same as the ones used in the corresponding SR, ensuring the FNP lepton processes are similar in the two regions.
In some cases, kinematic requirements are relaxed to increase the event yields.
The contribution of FNP background in the VR-SS regions is typically above 90$\%$, with the remaining background originating from diboson processes with two prompt leptons of the same charge.

\subsection{Control and validation regions for cut-and-count approach}

In total, two CRs and three VRs are defined for the cut-and-count-based approach, which are summarised in Table~\ref{tab:CCCRVRCuts}.
All regions are orthogonal to each other.
\CCCRTop and \CCCRVV are defined to constrain the top and diboson backgrounds, and are orthogonal to the SRs due to a $\nbjet^{20} >0 $ and \mtleptwo requirement, respectively. No strong dependence of the top and diboson modelling on \mtlepone was found, motivating common CRs for both \SRhigh and \SRlow.
The SM predictions are validated in the regions \CCVRDFHigh and \CCVRDFLow, which share the same kinematic requirements as the respective SRs, but include a DF requirement on the leptons.
In addition, regions using SS leptons are defined to validate the FNP lepton estimate in the cut-and-count phase space. To verify the estimate individually for electrons and muons, these regions are split by the flavour of the subleading lepton into \CCVRSSe and \CCVRSSm, where the leptons are ordered by \pt.

% The next lines are included from the .//tables/CCCRVRDefinitions.tex input file
\begin{table}[tbp]
\centering
\renewcommand{\arraystretch}{1.1}
\caption[]{Region definitions for CRs and VRs for the cut-and-count-based approach. Preselection requirements from Table \ref{tab:CommonCuts} are also applied.}
\adjustbox{max width=\textwidth}{
\begin{tabular}{ccccccc}
\toprule
\toprule
Variable                 & \CCCRVV         & \CCCRTop         & \CCVRSSe      & \CCVRSSm      & \CCVRDFHigh      & \CCVRDFLow       \\
\hline
Lepton charge and flavour  & OS              & OS              & SS $ee$ or $\mu e$ &    SS $e\mu$ or $\mu\mu$            & DFOS            & DFOS            \\
$\nbjet^{20}$            & $=0$           & $> 0$           & $=0$         & $=0$          & $=0$            & $=0$            \\
$\met$                   & --              & --               & --            & --             & $>300\ \GeV$    & $>300\ \GeV$    \\
\mtlepone        & $>100\ \GeV$   & $>100\ \GeV$    & --            & --             & $>100\ \GeV$    & $<100\ \GeV$    \\
\mtleptwo        & $<100\ \GeV$   & $>100\ \GeV$    & --            & --             & $>100\ \GeV$    & $>100\ \GeV$    \\
\bottomrule
\bottomrule
\end{tabular}
}
\label{tab:CCCRVRCuts}
\end{table}

% End of text imported from the .//tables/CCCRVRDefinitions.tex input file

\FloatBarrier

\subsection{Control and validation regions for BDT approach}

For the BDT approach, two CRs and four VRs per BDT, orthogonal to one another, are defined in Table \ref{tab:BDTCRVRDefintions}.
Kinematic selections, along with BDT score requirements, are applied to avoid overlap with the SRs, and to target specific background processes.
\BDTCRTop targets background events involving top quarks and is orthogonal to the SRs due to a $\nbjet^{20}>0$ requirement.
\BDTCRVV targets $VV$ background events and is orthogonal to the SRs through requirements on the BDT score and lepton flavour.
For $\BDTCRVV_{5+10}$ and $\BDTCRVV_{20}$, the requirement $\pt^{\ell_{2}}>7.5~\GeV$ is introduced to reduce the FNP-lepton contamination and increase the purity of the $VV$ background.
The naming convention of the BDT regions follows: $\mathrm{Region \mhyphen BDT_{X}^{Y}}$, where X indicates the BDT used to define the region and Y corresponds to an additional lepton flavour requirement, if any.
For example, the diboson CR based on the $\dmBase_{5+10}$ BDT is denoted by $\BDTCRVV_{5+10}$.

To validate the background normalisations extracted from these CRs, a set of validation regions is defined.
\BDTVRDF is orthogonal to the SRs through a DF lepton requirement while keeping all other selections of the corresponding SRs.
For the $\dmBase_{5+10}$ BDT this VR is slightly modified and the region is further split into $\BDTVRDF_{5+10}^{\mathrm{High}}$, which selects events with $\pt^{\ell_{2}} >10 \ \GeV$, and $\BDTVRDF_{5+10}^{\mathrm{Low}}$, which selects events with $\pt^{\ell_{2}} <10 \ \GeV$. The former VR is enriched in top and diboson events while the latter VR is dominated by FNP background, hence these VRs verify the modelling of these background components individually.
\BDTVRDFTop is equivalent to the VR-DF but with a $\nbjet^{20}>0$ requirement to validate background processes involving top quarks.
\BDTVRZZ is used to validate the modelling of the $ZZ/WZ$ backgrounds.
It is defined by requiring $|m_{\ell\ell}-m_{Z}|<10~\GeV$ in order to increase the contribution from events containing on-shell $Z$ bosons.
The FNP lepton estimate is validated in \BDTVRSS, which is defined by requiring a pair of SS leptons.
For the $\dmBase_{5+10}$ BDT selection the strategy of the FNP validation is refined, and the SS selection split according to the subleading lepton flavour into two VRs, $\BDTVRSS^{e}_{5+10}$ and $\BDTVRSS^{\mu}_{5+10}$. This split independently validates the FNP background estimate for subleading electrons and muons in the $\dmBase_{5+10}$ selection, where the SRs contain a significant FNP lepton component.
%

% The next lines are included from the .//tables/BDTCRVRDefintions.tex input file
\begin{table}[tbp]
\centering
\renewcommand{\arraystretch}{1.1}
\caption[]{Summary of region definitions for all CRs and VRs in the BDT approach. Only BDT score requirements and selections that enforce orthogonality between regions are shown. Preselection requirements from Table \ref{tab:CommonCuts} are also applied. A BDT score requirement that is exclusive of the upper bound is indicated by the use of ``)'' in defining the BDT score requirements.}
\adjustbox{max width=\textwidth}{
\begin{tabular}{c|c|c|c}
\toprule
\toprule
Region & BDT score & Additional requirements & Lepton flavour and charge \\
\hline
\hline
\multirow{5}{*}{\BDTCRVV} & $\mathrm{BDT}_{5+10} = [0.5, 0.93)$ & \multirow{2}{*}{$\pt^{\ell_{2}} > 7.5~\GeV$} & \multirow{5}{*}{OS leptons} \\
\cline{2-2}
& $\mathrm{BDT}_{20} = [0.7, 0.9)$ & & \\
\cline{2-2} \cline{3-3}
& $\mathrm{BDT}_{30} = [0.7, 0.88)$ & \multirow{3}{*}{---} & \\
\cline{2-2}
& $\mathrm{BDT}_{40+50} = [0.7, 0.9)$ & & \\
\cline{2-2}
& $\mathrm{BDT}_{60+75} = [0.7, 0.9)$ & & \\
\hline
\hline
\multirow{5}{*}{\BDTCRTop} & $\mathrm{BDT}_{5+10} = [0.7, 1.0]$ & \multirow{5}{*}{$\nbjet^{20} > 0$} & \multirow{5}{*}{SFOS leptons} \\
\cline{2-2}
& $\mathrm{BDT}_{20} = [0.7, 1.0]$ & & \\
\cline{2-2}
& $\mathrm{BDT}_{30} = [0.7, 1.0]$ & & \\
\cline{2-2}
& $\mathrm{BDT}_{40+50} = [0.7, 1.0]$ & & \\
\cline{2-2}
& $\mathrm{BDT}_{60+75} = [0.7, 1.0]$ & & \\
\hline
\hline
\multirow{5}{*}{\BDTVRDF} & $\mathrm{BDT}_{5+10} = [0.93, 1.0]$ & \multirow{5}{*}{---} & \multirow{5}{*}{DFOS leptons} \\
\cline{2-2}
& $\mathrm{BDT}_{20} = [0.9, 1.0]$ & & \\
\cline{2-2}
& $\mathrm{BDT}_{30} = [0.88, 1.0]$ & & \\
\cline{2-2}
& $\mathrm{BDT}_{40+50} = [0.9, 1.0]$ & & \\
\cline{2-2}
& $\mathrm{BDT}_{60+75} = [0.9, 1.0]$ & & \\
\hline
\hline
\multirow{5}{*}{\BDTVRDFTop} & $\mathrm{BDT}_{5+10} = [0.93, 1.0]$ & \multirow{5}{*}{$\nbjet^{20} > 0$} & \multirow{5}{*}{DFOS leptons} \\
\cline{2-2}
& $\mathrm{BDT}_{20} = [0.9, 1.0]$ & & \\
\cline{2-2}
& $\mathrm{BDT}_{30} = [0.88, 1.0]$ & & \\
\cline{2-2}
& $\mathrm{BDT}_{40+50} = [0.9, 1.0]$ & & \\
\cline{2-2}
& $\mathrm{BDT}_{60+75} = [0.9, 1.0]$ & & \\
\hline
\hline
\multirow{5}{*}{\BDTVRZZ} & $\mathrm{BDT}_{5+10} = [0.5, 1.0]$ & \multirow{5}{*}{$|m_{\ell\ell} - m_{Z}| < 10~\GeV$} & \multirow{5}{*}{SFOS leptons} \\
\cline{2-2}
& $\mathrm{BDT}_{20} = [0.7, 1.0]$ & & \\
\cline{2-2}
& $\mathrm{BDT}_{30} = [0.7, 1.0]$ & & \\
\cline{2-2}
& $\mathrm{BDT}_{40+50} = [0.7, 1.0]$ & & \\
\cline{2-2}
& $\mathrm{BDT}_{60+75} = [0.9, 1.0]$ & & \\
\hline
\hline
\multirow{5}{*}{\BDTVRSS} & $\mathrm{BDT}_{5+10} = [0.93, 1.0]$ & \multirow{1}{*}{Split by subleading $e$ or $\mu$} & \multirow{5}{*}{SS leptons} \\
\cline{2-2} \cline{3-3}
& $\mathrm{BDT}_{20} = [0.7, 1.0]$ & \multirow{4}{*}{---} & \\
\cline{2-2}
& $\mathrm{BDT}_{30} = [0.7, 1.0]$ & & \\
\cline{2-2}
& $\mathrm{BDT}_{40+50} = [0.7, 1.0]$ & & \\
\cline{2-2}
& $\mathrm{BDT}_{60+75} = [0.7, 1.0]$ & & \\
\bottomrule
\bottomrule
\end{tabular}
}
\label{tab:BDTCRVRDefintions}
\end{table}

% End of text imported from the .//tables/BDTCRVRDefintions.tex input file

% End of text imported from the .//sleptons/bkgEstimation.tex input file

%
\section{Systematic uncertainties}
%

% The next lines are included from the .//sleptons/systematics.tex input file
The SM and signal predictions used in the cut-and-count and BDT-based selections are affected by systematic uncertainties that are due to both experimental effects and uncertainties in MC generator modelling.
The $VV$ and top backgrounds are normalised in dedicated CRs, hence systematic uncertainties in these backgrounds only affect the extrapolation of these predictions from the CRs to the VRs and SRs. A summary of all systematic effects for the cut-and-count and BDT-based SRs is shown in Figures~\ref{fig:CCSysSummary} and~\ref{fig:BDTSysSummary} respectively. The same set of systematic uncertainties is considered for the cut-and-count and BDT-based selections.

\begin{figure}[tbp]
\centering
\subfloat[]{
\includegraphics[width=0.9\textwidth]{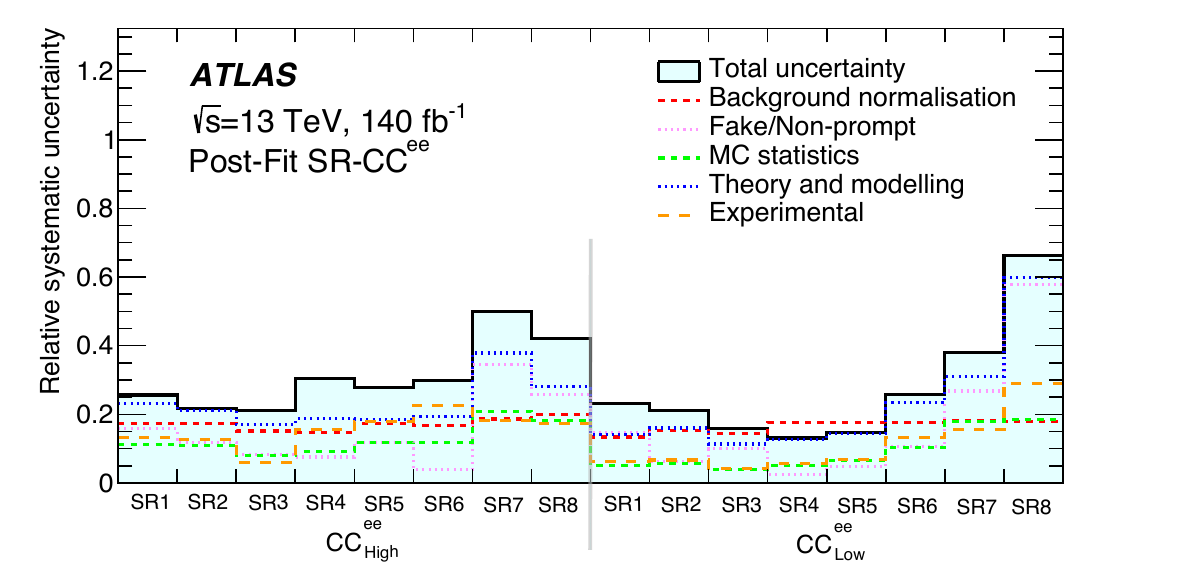}
}\\
\subfloat[]{
\includegraphics[width=0.9\textwidth]{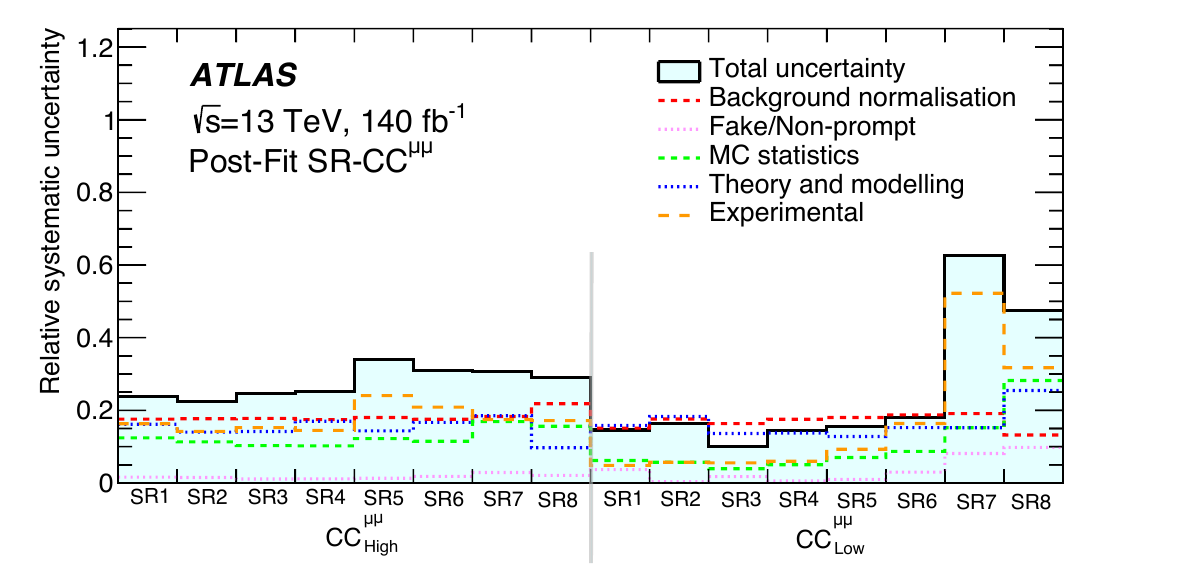}
}
\hfil
\caption{Relative systematic uncertainties in the post-fit SM background estimates in the cut-and-count-based (a) di-electron and (b) di-muon SRs obtained from a background-only fit to the CRs.
The ``Fake/Non-Prompt'' category reflects uncertainties impacting the FNP background estimate.
Uncertainties originating from the limited size of the MC samples used to model the irreducible background contributions are contained in the ``MC statistics'' category.
The ``Background normalisation'' category reflects uncertainty in normalisation factors for $VV$ and top backgrounds extracted from \CCCRVV and \CCCRTop.
The ``Theory and modelling'' category includes the different sources of theoretical modelling uncertainties for the top and diboson backgrounds.
The ``Experimental'' category covers detector related uncertainties from the reconstruction and selection of objects in the analysis.
The breakdown into individual categories is derived by iteratively performing fits where only the parameters associated with a particular category are allowed to float with all other parameters set constant. The uncertainties are then propagated to the background predictions using the covariance matrix of the nominal fit where all parameters are allowed to float~\cite{Baak:2014wma}.
The individual uncertainties are correlated and do not necessarily add up in quadrature to the total uncertainty.
}

\label{fig:CCSysSummary}
\end{figure}

\begin{figure}[tbp]
\centering
\subfloat[]{
\includegraphics[width=0.9\textwidth]{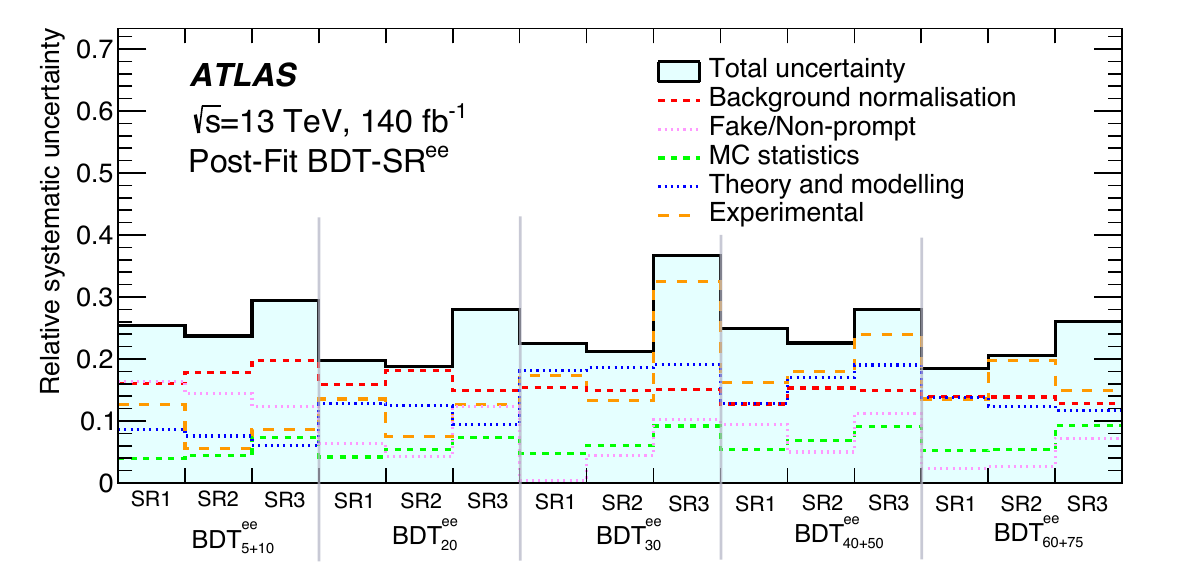}
}\\
\subfloat[]{
\includegraphics[width=0.9\textwidth]{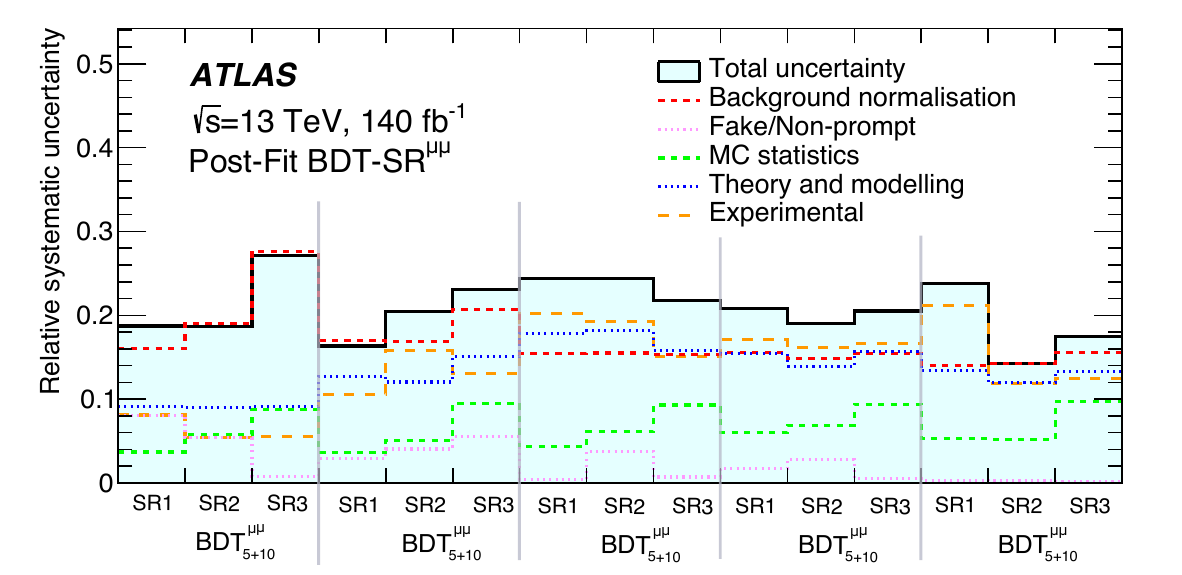}
}
\caption{Relative systematic uncertainties in the post-fit SM background estimates in the BDT-based (a) di-electron and (b) di-muon SRs obtained from a background-only fit to the CRs.
The ``Fake/Non-Prompt'' category reflects uncertainties impacting the FNP background estimate.
Uncertainties originating from the limited size of the MC samples used to model the irreducible background contributions are contained in the ``MC statistics'' category.
The ``Background normalisation'' category reflects uncertainty in normalisation factors for $VV$ and top backgrounds extracted from \BDTCRTop and \BDTCRVV.
The ``Theory and modelling'' category includes the different sources of theoretical modelling uncertainties for the top and diboson backgrounds.
The ``Experimental'' category covers detector related uncertainties from the reconstruction and selection of objects in the analysis.
The breakdown into individual categories is derived by iteratively performing fits where only the parameters associated with a particular category are allowed to float with all other parameters set constant. The uncertainties are then propagated to the background predictions using the covariance matrix of the nominal fit where all parameters are allowed to float~\cite{Baak:2014wma}.
The individual uncertainties are correlated and do not necessarily add up in quadrature to the total uncertainty.}
\label{fig:BDTSysSummary}
\end{figure}

Background and signal distributions predicted by MC samples are subject to a set of uncertainties in the physics objects used in the analysis.
Uncertainties in electrons~\cite{EGAM-2018-01} and muons~\cite{PERF-2015-10} arise from uncertainties in the corrections applied in MC simulation of their reconstruction, identification, and isolation efficiencies, as well as in the corrections to their momentum resolution and energy scale.
The jet energy resolution and scale uncertainties are applied as a function of the $\pt$ and $\eta$ of the jet and the uncertainties are derived using a combination of data and MC samples, using measurements of for example the jet \pt balance in dijet, $Z+$~jet and $\gamma+$~jet events ~\cite{JETM-2018-05,PERF-2014-02}. Further jet-related uncertainties originate from efficiency corrections for tagging pile-up jets~\cite{PERF-2014-03} and the identification of $b$-jets~\cite{FTAG-2018-01,FTAG-2020-08}.
The \met uncertainties are estimated by propagating the uncertainties in the energy and momentum scale of each of the objects entering the calculation of \met, and include the uncertainties in the soft-term resolution and scale~\cite{PERF-2016-07}.
Experimental uncertainties are generally implemented as two-sided variations relative to the nominal event yields provided by the simulation.

The uncertainty in the FNP background estimate based on the fake-factor method has several components.
Uncertainties in the fake factors originate from the size of the data sample in which these are derived and further kinematic dependencies than the ones already considered.
The size of the latter uncertainty is 40\% for electrons and 10\% for muons.
Uncertainties in the subtraction of the prompt-lepton contamination were estimated by varying their size and are found to be small compared with the other uncertainties.
Additional statistical uncertainties arise from the limited size of the loose-lepton control samples.
If the control sample does not contain any events, an additional one-sided uncertainty is applied based on the largest fake factor applicable to provide an upper limit on the FNP contribution.
Finally, a non-closure uncertainty is derived to account for residual discrepancies observed in the SS VRs that are not fully covered by the uncertainties described above.
This uncertainty is evaluated separately for electrons and muons and is parameterised as a function of the lepton \pt with sizes ranging from 5\% to 100\%.

Uncertainties related to MC generator modelling of the diboson and top backgrounds arise in particular from the choice of the QCD factorisation and renormalisation scales.
These are assessed by varying the associated generator parameters up and down by a factor of two around their nominal values, avoiding combinations that differ by a factor of 4.
Furthermore, uncertainties associated with the choice of PDF set and on the strong coupling constant \alphas are also considered.
Additional uncertainties are evaluated on the top background.
The impact of using a different parton shower and hadronisation model was evaluated by comparing the nominal top samples generated with \POWHEGBOX to samples that were interfaced to the \HERWIG[]~\cite{Bahr:2008pv,Bellm:2015jjp} instead of \PYTHIA[].
The uncertainty associated with the matching between the matrix element and the parton shower is estimated by a variation of the parameter that regulates the definition of the vetoed region of the showering~\cite{Hoche:2021mkv}.
These variations are evaluated for both the \ttbar and single-top processes and are treated as correlated within the top background.
A further uncertainty is considered on the treatment of the interference between the $\ttbar$ and $Wt$ processes~\cite{Frixione:2008yi,ATL-PHYS-PUB-2016-020}.
The uncertainty is assessed by comparing the nominal $Wt$ sample that uses the diagram removal scheme~\cite{Frixione:2008yi} to an alternative sample employing the diagram subtraction scheme~\cite{Frixione:2008yi,ATL-PHYS-PUB-2016-020}.
For the residual, irreducible backgrounds estimated directly from simulation, an overall $50\%$ ad-hoc normalisation uncertainty is applied to take into account uncertainties in their cross-sections.
As these backgrounds have only a very minor contribution, the impact of this latter uncertainty in the SM predictions in the SRs is small.

Overall, the leading experimental uncertainties in the background estimates were found to originate from the jet energy resolution as well as the resolution and scale of the \met soft term.
Uncertainties in the lepton efficiency corrections are small compared to the other experimental uncertainties.
The dominant MC-modelling uncertainties are the QCD scale variations and those due to the matching between matrix element and parton shower for the top background.
The overall uncertainty in SRs in the cut-and-count-based approach ranges from 20--50\% in the di-electron SRs and 20--40\% in the di-muon SRs.
Uncertainties increase in particular in the tail of the $\mttwo^{100}$ distributions, where statistical accuracy of the MC predictions is reduced.
The contribution of the FNP lepton related uncertainties is negligible in the cut-and-count-based SRs.
For the BDT-based approach, the overall uncertainty ranges from 20--35\% in the di-electron SRs and 15--25\% in the di-muon SRs.
The leading uncertainties vary across the BDT selections as the background composition changes, but their overall magnitude remains similar.
In the \VLowSR, the normalisation factor uncertainties are dominant due to the large contribution from the FNP lepton background in the CRs associated with this BDT.
These uncertainties are reduced in the higher mass-splitting BDT regions, where the FNP contribution to the corresponding CRs is lower.

In addition to the experimental uncertainties described above, the predictions of the benchmark signals for slepton-pair production are also subject to a set of theoretical uncertainties, including cross-section and shape uncertainties due to the QCD renormalisation and factorisation and the choice of PDF set. These are accessed following the same procedure that is used for the diboson and top backgrounds. Only a mild dependence on \mslepton and \dm was found with average values of about 20--30\% and 3\% associated with the scale and PDF variations, respectively. Uncertainties in the value of \alphas were found to be negligible for the signal processes.

%
%
%
%

% End of text imported from the .//sleptons/systematics.tex input file

%
\section{Results}
%

% The next lines are included from the .//sleptons/results.tex input file
Observed data in the CRs, VRs and SRs are compared with the SM predictions using a profile likelihood method~\cite{Cowan:2010js} implemented in the HistFitter package~\cite{Baak:2014wma}.
Systematic uncertainties are incorporated as nuisance parameters with Gaussian constraints within the likelihood, where experimental systematic uncertainties are correlated between signal and background across all regions.

Several different fit configurations are used to derive the results presented below and are all carried out individually for the cut-and-count-based and each of the BDT selections.
The first configuration, referred to as CR-only fit, represents a background-only fit to the two CRs to extract the normalisation factors $\mu_{\text{VV}}$ and $\mu_{\text{top}}$ of the diboson and top backgrounds, respectively.
The fit results can then be extrapolated to the VRs and SRs to compare the post-fit SM predictions with the observed data.
The second fit configuration, referred to as discovery fit, includes both CRs and exactly one SR bin of \SRDisc.
This configuration is used to derive a $p$-value representing the probability that the data in the SR bin are compatible with the background-only hypothesis.
In the third fit configuration, a group of CRs and SRs is fitted simultaneously to determine the probability that the observed data are compatible with a specific signal model, with such fits consequently being referred to as exclusion fits.
These are carried out individually for the cut-and-count-based and each of the BDT-based selections.

\subsection{Results for control and validation regions}

The normalisation factors derived by a CR-only fit in the cut-and-count and BDT-based approaches are summarised in Table~\ref{tab:NormFactors} and visualised along with the data and the pre-fit CR predicted event yields in Figure~\ref{fig:CRSummary}. The mild differences between the diboson and top normalisation factors among the various BDTs originate from the different phase spaces covered by each BDT selection.

Figures~\ref{fig:VRSummaryCC} and~\ref{fig:VRSummaryBDT} show the data and the post-fit SM predictions in the VR-DF regions of the cut-and-count and all VRs of the BDT selections, respectively. Good agreement of the predictions with the observed data is found in all VRs, where all deviations are found to be below two standard deviations as approximated with the prescription from Ref.~\cite{ATL-PHYS-PUB-2020-025}.
The largest discrepancy is in $\BDTVRZZ_{60+75}$, which was found to result from the $VV$ normalisation factor derived in $\BDTCRVV_{60+75}$ that predominantly contains $W^+W^-$ events and therefore does not extrapolate as well to the higher proportion of on-shell $ZZ$ decays in this VR\@.
The SM predictions were found to agree with the observed kinematic distributions in \CCVRSSe, \CCVRSSm,  $\BDTVRSS^e_{5+10}$ and $\BDTVRSS^\mu_{5+10}$ within uncertainties, verifying the validity of the data-driven FNP estimate.
%

%
%
%

%
%

% The next lines are included from the .//tables/NormFactors.tex input file
\begin{table}[tbp]
\centering
\renewcommand{\arraystretch}{1.1}
\caption[]{Normalisation factors $\mu_{\text{VV}}$ and $\mu_{\text{Top}}$ for the $VV$ and top backgrounds, respectively, extracted from CR-only fits using the cut-and-count and BDT-based selections. The associated uncertainties include all statistical and systematic contributions.}
\begin{tabular}[h]{lcc}
\toprule
\toprule
Normalisation factor & $\mu_{\text{VV}}$  & $\mu_{\text{Top}}$\\
\midrule
CC & $0.97 \pm 0.23$ &  $0.90 \pm 0.24$ \\
\midrule
$\mathrm{BDT}_{5+10}$ & $0.88 \pm 0.36$ &  $0.79 \pm 0.24$ \\
$\mathrm{BDT}_{20}$ & $0.89 \pm 0.23$ &  $0.95 \pm 0.27$ \\
$\mathrm{BDT}_{30}$ & $0.94 \pm 0.18$ &  $0.93 \pm 0.27$ \\
$\mathrm{BDT}_{40+50}$ & $0.98 \pm 0.18$ &  $0.82 \pm 0.23$ \\
$\mathrm{BDT}_{60+75}$ & $1.19 \pm 0.20$ &  $0.82 \pm 0.22$ \\
\bottomrule
\bottomrule
\end{tabular}
\label{tab:NormFactors}
\end{table}

% End of text imported from the .//tables/NormFactors.tex input file

\begin{figure}[tbp]
\centering
\centering\includegraphics[width=1.0\textwidth]{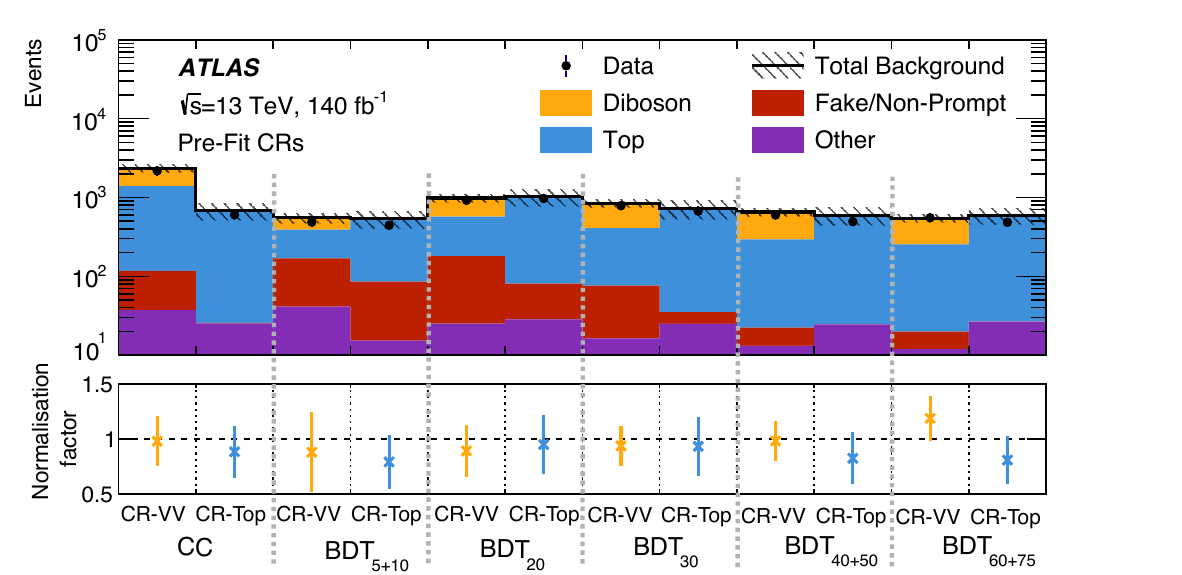}
\caption{Data (dots) and pre-fit expected (histograms) event yields in the CRs for both the cut-and-count and BDT-based approaches.
Shown in the lower panel are the normalisation factors (crosses), for $\mu_{\text{VV}}$ (orange) and $\mu_{\text{Top}}$ (blue) derived from a CR-only fit, along with the associated uncertainties.
Uncertainties in the background estimates include both statistical and systematic uncertainties.}
\label{fig:CRSummary}
\end{figure}

\begin{figure}[tbp]
\centering
\includegraphics[width=0.95\textwidth]{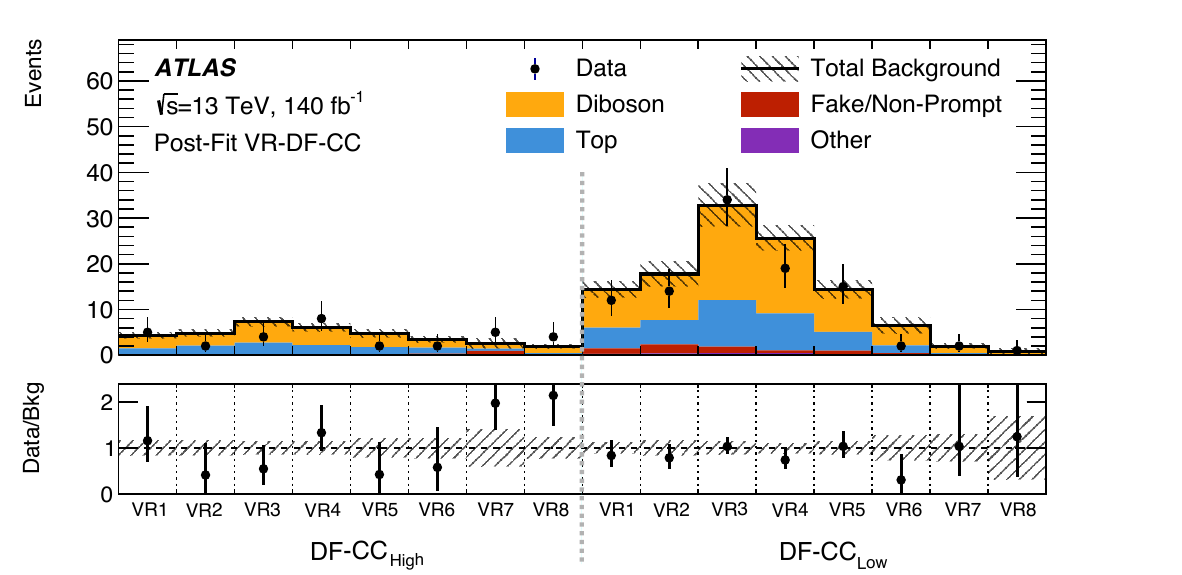}
\caption{Data (dots) and post-fit SM predictions (histograms) in \CCVRDFHigh and \CCVRDFLow for the cut-and-count approach using a CR-only fit.
The lower panel shows the ratio of observed data to the total post-fit SM prediction.
Uncertainties in the background estimates include both statistical and systematic uncertainties.}
\label{fig:VRSummaryCC}
\end{figure}

\begin{figure}[tbp]
\centering
\includegraphics[width=0.95\textwidth]{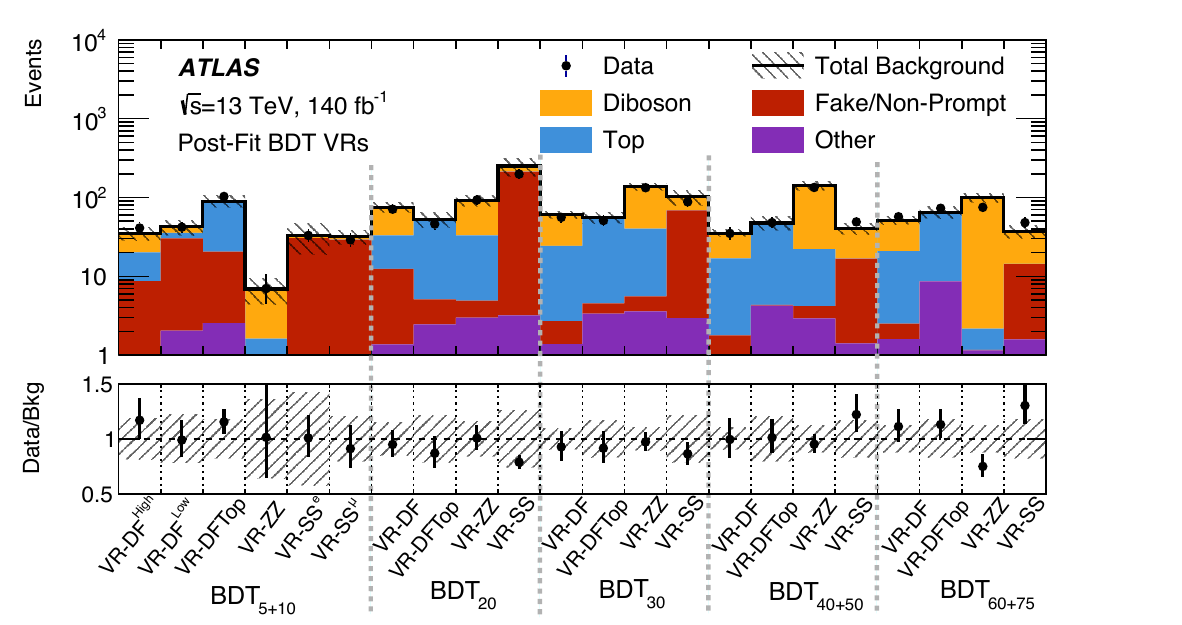}
\caption{Data (dots) and post-fit SM predictions (histograms) in the BDT-based VRs using a CR-only fit.
The lower panel shows the ratio of observed data to the total post-fit SM prediction.
Uncertainties in the background estimates include both statistical and systematic uncertainties.}
\label{fig:VRSummaryBDT}
\end{figure}

\subsection{Results for cut-and-count signal regions}

The observed and predicted event yields of the \SRhighe, \SRlowe, \SRhighm and \SRlowm selections are shown in Figure~\ref{fig:CCSRSummary}.
The SM predictions agree well with the data within their uncertainties and no substantial excess of events is found.
The most significant deviation is in \SRlowThreeee, which covers the $\mttwo^{100}$ range of 110--120~\GeV, where the observed yield is significantly smaller than the prediction. In the corresponding $\mathrm{VR3\mhyphen DF \mhyphen CC_{\mathrm{Low}}}$ and \SRlowThreemm regions, which contain similar top and background events due to their lepton-flavour symmetry, good agreement is observed.
Hence, this is considered as a statistical fluctuation.
All CRs and SRs used in the BDT-based analysis were found to have only minor overlap with the phase space covered by this bin and show no evidence of a similar deviation.

\begin{figure}[tbp]
\centering
\subfloat[]{
\includegraphics[width=\textwidth]{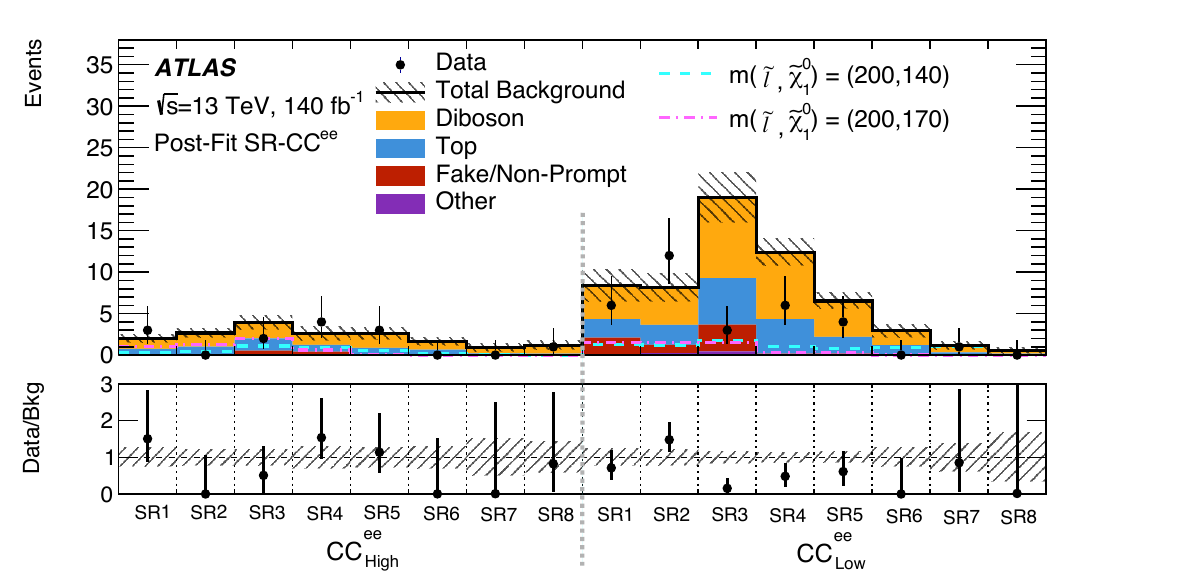}
}\\
\subfloat[]{
\includegraphics[width=\textwidth]{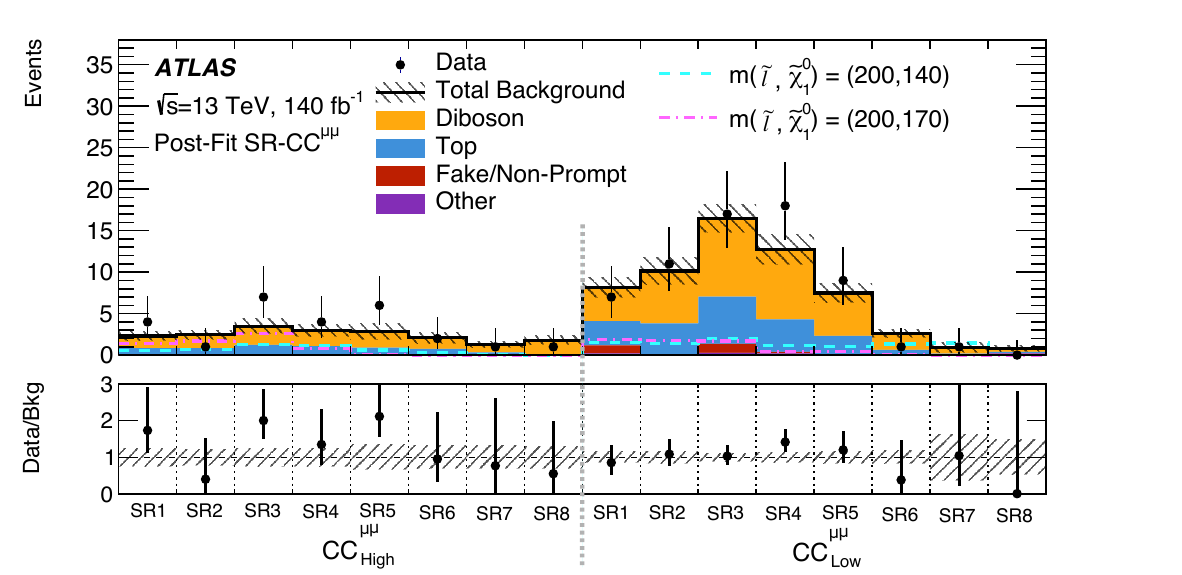}
}\\
\caption{Data (dots) and post-fit SM predictions (histograms) in (a)
\SRhighe and \SRlowe, and (b) \SRhighm and \SRlowm. The predictions are obtained from a CR-only fit.
Uncertainties in the background estimates include both statistical and systematic uncertainties.
The lower panel shows the ratio of observed data to the total post-fit SM prediction.
The dashed lines show predicted yields for two benchmark signal models corresponding to $m(\slep, \chioz) = (200, 140)~\GeV$ and $(200, 170)~\GeV$, assuming $m(\slepL) = m(\slepR)$.}
\label{fig:CCSRSummary}
\end{figure}

Observed and predicted event yields in the lepton-flavour inclusive \SRDisc are shown in Figure~\ref{fig:CCInclusive}, which are used to test for the presence of excesses.
A discovery-fit configuration is formed with each of the \SRDisc that contains an additional signal model with an unconstrained signal-strength parameter in the SR under study to estimate the size of any BSM contribution still compatible with the observed data.
The background normalisation factors $\mu_{\text{VV}}$ and $\mu_{\text{Top}}$ are treated as free parameters in the discovery fit while any signal contamination in the CRs is neglected.
The \cls prescription~\cite{myRead:2002hq} is used to set upper limits at the 95\% confidence level (CL) on the signal cross-sections, $\sigma_{\text{obs}}$, which is defined as the product of cross-section, acceptance, and efficiency for BSM processes.
The results of these fits, carried out with 5000 pseudo-experiments for each SR, are presented in Table~\ref{tab:CCSRSFUpperLimits}.
Generic non-SM processes that predict more than 9.1 events in $\SRDisc^{<105}$ are excluded at 95\% CL, while this number increases to 18.0 events when considering $\SRDisc^{<180}$, which covers the complete $\mttwo^{100}$ spectrum considered in this analysis.
This corresponds to upper limits on the visible cross-sections of 0.065~fb and 0.129~fb, respectively.

% The next lines are included from the .//tables/CCSRSFUpperLimit.tex input file

\begin{table}[tbp]
\centering
\renewcommand{\arraystretch}{1.1}
\caption{Results of the fits in the single-bin inclusive \SRDisc, where the first column indicates the SR under study. The expected total SM background event yields, $N_{
\text{exp}}$,  and the observed event yields, $N_{\text{obs}}$, are shown in the second and third columns of the table.
The SM expectation is calculated using the discovery fit configuration, where each SR individually is used in a simultaneous fit with the CRs.
The following columns show the observed 95\% CL upper limits  on the cross-section, $\langle \epsilon \sigma \rangle^{95}_{\text{obs}}$, and on
the observed and expected number of new-physics events, denoted $S^{95}_{\text{obs}}$ and $S^{95}_{\text{exp}}$, respectively.
The $\pm 1\sigma$ variations of $S^{95}_{\text{exp}}$ are also shown.
The final columns show the $p$-value for the SM-only hypothesis, $p(S=0)$, and the corresponding Gaussian-equivalent statistical significance, where $p$-values are capped at $0.5$.
The $p$-values, significances are calculated using individual discovery fits, with 5,000 pseudo-experiments.
Cases where the total SM expectation is greater than the observed data are indicated by a dash, `$\text{--}$', and no $p$-value or significance is reported.}
\label{tab:CCSRSFUpperLimits}
\begin{tabular}[h]{lcccccccc}
\toprule
\toprule
Region & $N_{\text{exp}}$ & $N_{\text{obs}}$ & $\langle \epsilon \sigma \rangle^{95}_{\text{obs}} [\text{fb}]$ & $S^{95}_{\text{obs}}$ & $S^{95}_{\text{exp}}$ & $p(S=0)$ & $\sigma$ \\
\hline
$\SRDisc^{< 105}$  & $4.8 \pm 0.9$  & $7$   & $0.065$ & $9.1$  & $6.2^{+2.2}_{-1.7}$  & $0.073$ & $1.45$ \\
$\SRDisc^{< 110}$  & $9.1 \pm 1.3$  & $8$   & $0.049$ & $6.9$  & $7.8^{+2.8}_{-2.2}$  & $\text{--}$ & $\text{--}$   \\
$\SRDisc^{< 120}$  & $16.9 \pm 1.8$ & $17$  & $0.074$ & $10.3$ & $10.1^{+3.8}_{-2.8}$  & 0.5 & 0  \\
$\SRDisc^{< 130}$  & $23.1 \pm 2.4$ & $25$  & $0.097$ & $13.6$ & $11.9^{+5.0}_{-3.2}$ & $0.35$ & $0.38$ \\
$\SRDisc^{< 140}$  & $29.8 \pm 2.9$ & $34$  & $0.131$ & $18.4$ & $14.4^{+5.7}_{-4.2}$ & $0.21$ & $0.79$ \\
$\SRDisc^{< 150}$  & $33.2 \pm 3.3$ & $36$  & $0.130$ & $18.2$ & $14.8^{+5.9}_{-3.7}$ & $0.30$ & $0.52$ \\
$\SRDisc^{< 160}$  & $35.0 \pm 3.5$ & $37$  & $0.125$ & $17.5$ & $15.6^{+6.4}_{-4.9}$ & $0.37$ & $0.32$ \\
$\SRDisc^{< 180}$  & $38 \pm 4$ & $39$  & $0.129$ & $18.0$ & $16.5^{+6.5}_{-3.7}$ & $0.40$ & $0.25$ \\

\bottomrule
\bottomrule
\end{tabular}
\end{table}

% End of text imported from the .//tables/CCSRSFUpperLimit.tex input file

\begin{figure}[tbp]
\centering
\includegraphics[width=\textwidth]{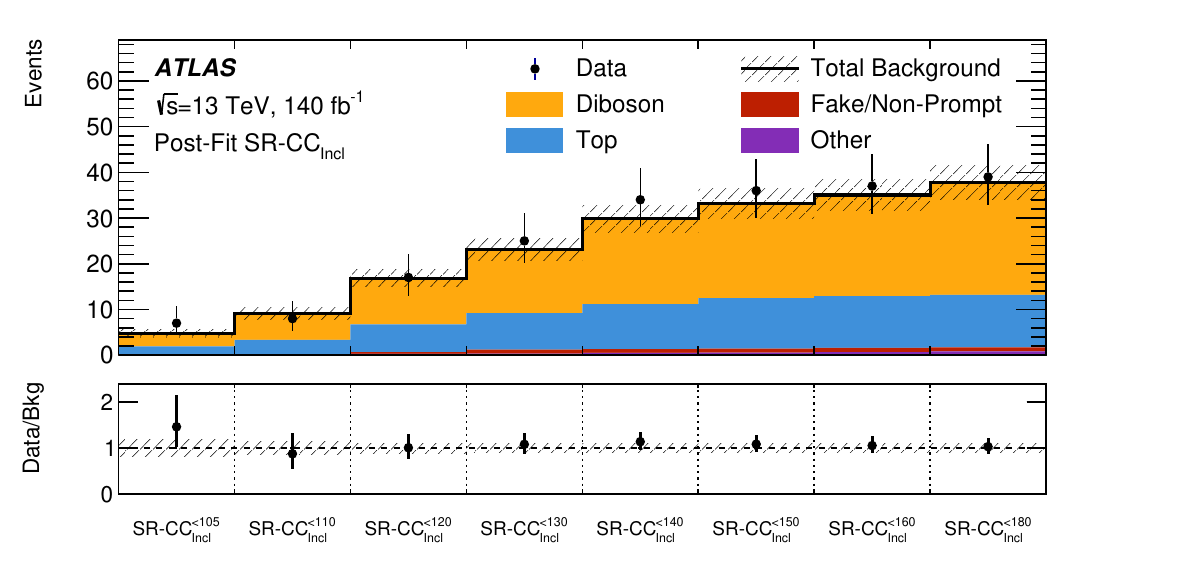}
\caption{Data (dots) and post-fit SM predictions (histograms) in the \SRDisc defined in the cut-and-count-based analysis where each bin includes the yields from the previous bin by construction.
The lower panel shows the ratio of observed data to the total post-fit SM prediction.}
\label{fig:CCInclusive}
\end{figure}

\FloatBarrier

\subsection{Results for BDT signal regions}

The observed data and predicted SM event yields for the BDT-based signal selections from a CR-only fit are shown in Figure~\ref{fig:BDTSRSummary}, along with the ratio of the observed data to the post-fit SM expectation.
While in most SRs the SM predictions agree well with the data, within uncertainties, a few bins show excesses of around two standard deviations, which are further discussed below.

Table \ref{tab:BDTSig} summarises the observed and total post-fit SM yields for each BDT SR, along with the statistical significance of the difference between observed data and the post-fit predicted yield for each SR\@.
The significances and post-fit yields are determined using a discovery fit, mirroring that used by the cut-and-count inclusive SRs, and using 15,000 pseudo-experiments.
The CRs are fit simultaneously with one SR at a time, allowing for the extraction of an individual single-bin significance for each SR.
\begin{figure}[tbp]
\centering
\subfloat[]{
\includegraphics[width=\textwidth]{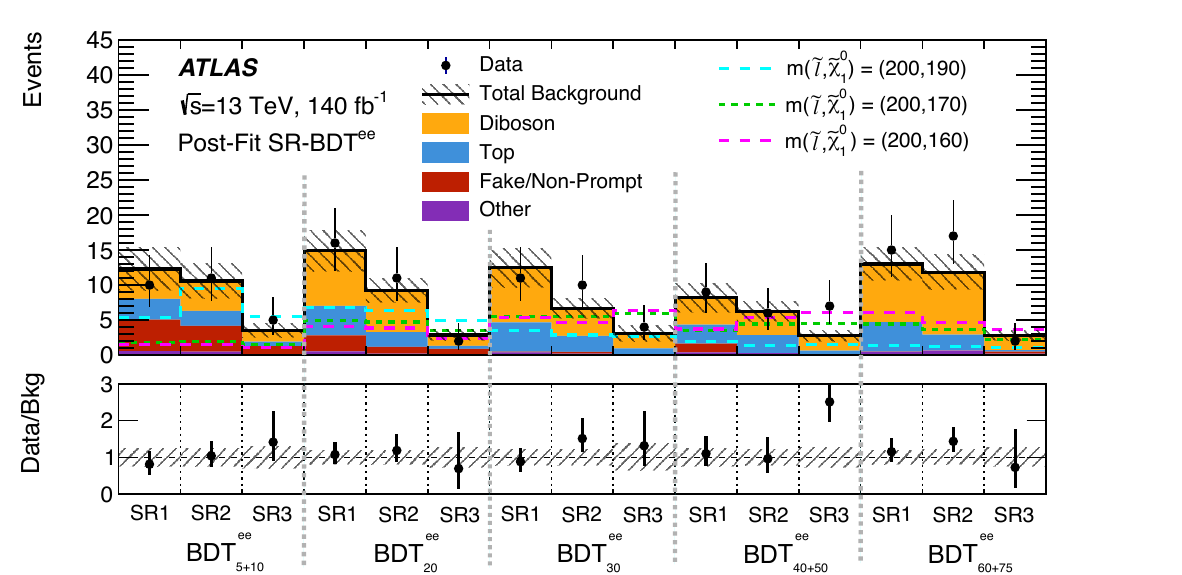}
}\\
\subfloat[]{
\includegraphics[width=\textwidth]{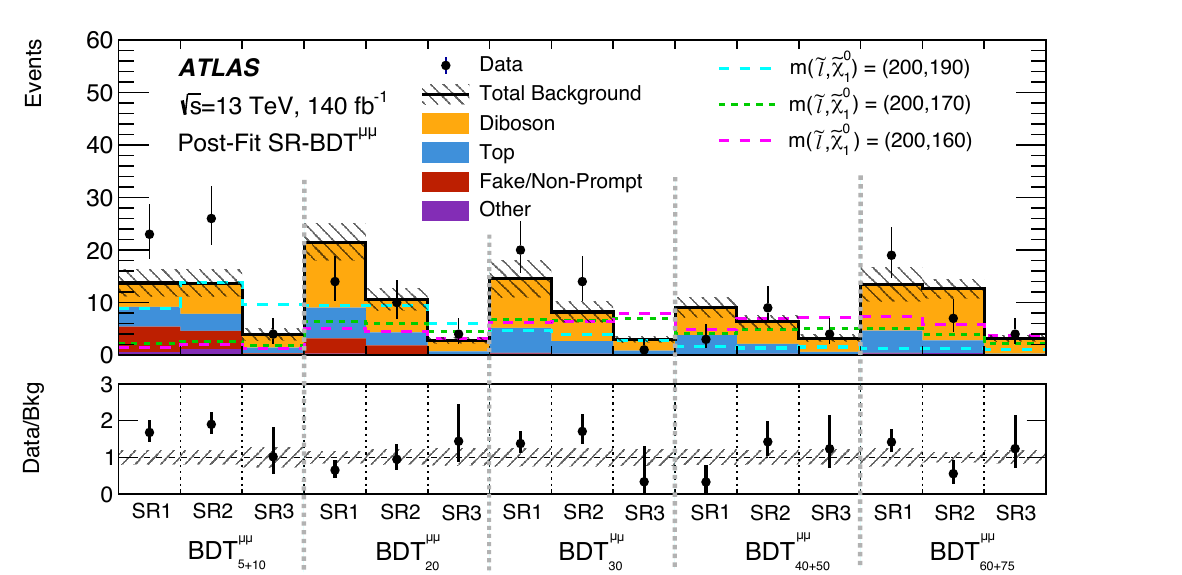}
}
\caption{Data (dots) and post-fit SM predictions (histograms) in (a) \BDTSRee{} and (b) \BDTSRmm{}. The SM predictions are obtained from a CR-only fit.
Uncertainties in the background estimates include both statistical and systematic uncertainties.
The bottom panel shows the ratio of observed data to the total post-fit SM prediction.
The dashed lines show predicted yields for three benchmark signal models corresponding to $m(\slep, \chioz) = (200, 190)~\GeV$, $(200, 170)~\GeV$ and $(200, 160)~\GeV$, assuming $m(\slepL) = m(\slepR)$.}
\label{fig:BDTSRSummary}
\end{figure}

% The next lines are included from the .//tables/BDTSignificance.tex input file
\begin{table}[tbp]
\centering
\renewcommand{\arraystretch}{1.25}
\caption{Observed and the total post-fit SM expectation, along with the statistical significance of their deviation, shown separately for (left) \BDTSRee and (right) \BDTSRmm.
The expected total SM background event yields, $N_{\text{exp}}$, and the observed event yields, $N_{\text{obs}}$, are shown in the second and third columns of the table.
The SM expectation is calculated using the CR-only fit configuration, where only the two CRs of the relevant BDT are fit simultaneously.
The fourth column shows the statistical significance of the deviation between the expected total SM prediction and observed data, calculated using individual discovery fits with 15,000 pseudo-experiments.
Cases where the expected total SM prediction is greater than the observed data are indicated by a dash, `$\text{--}$', and no significance is reported.}
\label{tab:BDTSig}
\begin{minipage}{0.48\linewidth}
\centering
\footnotesize
\setlength{\tabcolsep}{4pt}
\sisetup{table-align-text-post = false, table-format=3.2}
\begin{tabular}{l S[table-format=3.2, output-decimal-marker={.}] @{\,$\pm$} S[table-format=2.2, output-decimal-marker={.}] c S[table-format=2.2]}
\toprule
\toprule
Region ($ee$) & \multicolumn{2}{c}{$N_{\text{exp}}$} & $N_{\text{obs}}$ & $\sigma$\\
\midrule
\VLowSROnee & 12.3 & 3.1 & 10  & $\text{--}$  \\
\VLowSRTwoe & 10.6 & 2.5 & 11  &  0.17 \\
\VLowSRThreee & 3.5 & 1.0 & 5   &  0.59 \\
\midrule
\LowSROnee & 14.9 & 2.9 & 16  &  0.28 \\
\LowSRTwoe & 9.3 & 1.7 & 11  &  0.44 \\
\LowSRThreee & 2.9 & 0.8 & 2   & $\text{--}$ \\
\midrule
\MidSROnee & 12.5 & 2.8 & 11  & $\text{--}$ \\
\MidSRTwoe & 6.6 & 1.4 & 10  &  1.0 \\
\MidSRThreee & 3.0 & 1.1 & 4   &  0.53 \\
\midrule
\HighSROnee & 8.2 & 2.0 & 9   &  0.26 \\
\HighSRTwoe & 6.2 & 1.4 & 6   & $\text{--}$ \\
\HighSRThreee & 2.8 & 0.8 & 7   &  2.0 \\
\midrule
\VHighSROnee & 13.0 & 2.4 & 15  &  0.69 \\
\VHighSRTwoe & 11.9  & 2.4 & 17  &  1.2 \\
\VHighSRThreee & 2.8 & 0.7 & 2   & $\text{--}$ \\
\bottomrule
\bottomrule
\end{tabular}
\end{minipage}%
\hspace{1em}
\begin{minipage}{0.48\linewidth}
\centering
\footnotesize
\setlength{\tabcolsep}{4pt}
\sisetup{table-align-text-post = false, table-format=3.2}
\begin{tabular}{l S[table-format=3.2, output-decimal-marker={.}] @{\,$\pm$} S[table-format=2.2, output-decimal-marker={.}] c S[table-format=2.2]}
\toprule
\toprule
Region ($\mu\mu$) & \multicolumn{2}{c}{$N_{\text{exp}}$} & $N_{\text{obs}}$ & $\sigma$\\
\midrule
\VLowSROnem & 13.7 & 2.6 & 23  &  1.9 \\
\VLowSRTwom & 13.7 & 2.5 & 26  &  2.4 \\
\VLowSRThreem & 3.9 & 1.1 & 4   & $\text{--}$ \\
\midrule
\LowSROnem & 21.5 & 3.5 & 14  & $\text{--}$ \\
\LowSRTwom & 10.6 & 2.2 & 10  & $\text{--}$ \\
\LowSRThreem & 2.8 & 0.6 & 4   &  0.68 \\
\midrule
\MidSROnem & 14.6  & 3.5 & 20  &  1.1 \\
\MidSRTwom & 8.2 & 2.0 & 14  &  1.4 \\
\MidSRThreem & 3.0  & 0.6 & 1   & $\text{--}$ \\
\midrule
\HighSROnem & 9.1 & 1.9 & 3   & $\text{--}$ \\
\HighSRTwom & 6.3 & 1.2 & 9   &  1.0 \\
\HighSRThreem & 3.2 & 0.7 & 4   &  0.46 \\
\midrule
\VHighSROnem & 13.4  & 3.2 & 19  &  1.1 \\
\VHighSRTwom & 12.6 & 1.8 & 7   & $\text{--}$ \\
\VHighSRThreem & 3.2& 0.6 & 4   &  0.07 \\
\bottomrule
\bottomrule
\end{tabular}
\end{minipage}
\end{table}

% End of text imported from the .//tables/BDTSignificance.tex input file

In the di-electron channel the most discrepant region is \HighSRThreee, with an excess of events that corresponds to a local significance of $2.0\,\sigma$.
Excesses of smaller statistical significance are observed in \MidSRTwoe and \VHighSRTwoe. As these SRs are not statistically independent, the latter mild excesses are caused to a large extent by the same events that are also found in \HighSRThreee.
There is no substantial overlap of the events found in \HighSRThreee with the events that enter into any of the cut-and-count-based SRs.

In the di-muon channel the most discrepant SRs are \VLowSROnem and \VLowSRTwom, where the excesses have statistical significances of $1.9\,\sigma$ and $2.4 \,\sigma$, respectively.
By construction there is no overlap of events between these two regions.
For illustration, kinematic distributions of variables important in the training of this BDT are shown in Figure~\ref{fig:LowDMExcess}, inclusively for all three bins of $\BDTSRmm_{5+10}$.
Overlaid are also kinematic distributions of two example signal models: ($m_{\tilde{\mu}}, m_{\chioz}) = (150~\GeV, 140~\GeV)$ and $(200~\GeV, 190~\GeV)$.

The excess of events is accumulated at \dRll values of around 3.0, with no such feature being observed in the associated set of VRs or in the electron counterpart of this signal selection, $\BDTSRee_{5+10}$.
There is minimal overlap between events in the $\BDTSRmm_{5+10}$ regions and those in other BDT SRs or any of the cut-and-count-based SRs.
The $\dmBase_{5+10}$ BDT contains a large proportion of events with low-\pt leptons, which is a feature not present in any other SRs.
As a result, the largest data event overlap occurs with the signal regions of the neighbouring BDT, $\BDTSRmm_{20}$, which shares $15\%$ of events with $\BDTSRmm_{5+10}$.
\begin{figure}[tbp]
\centering
\subfloat[]{
\includegraphics[width=0.49\textwidth]{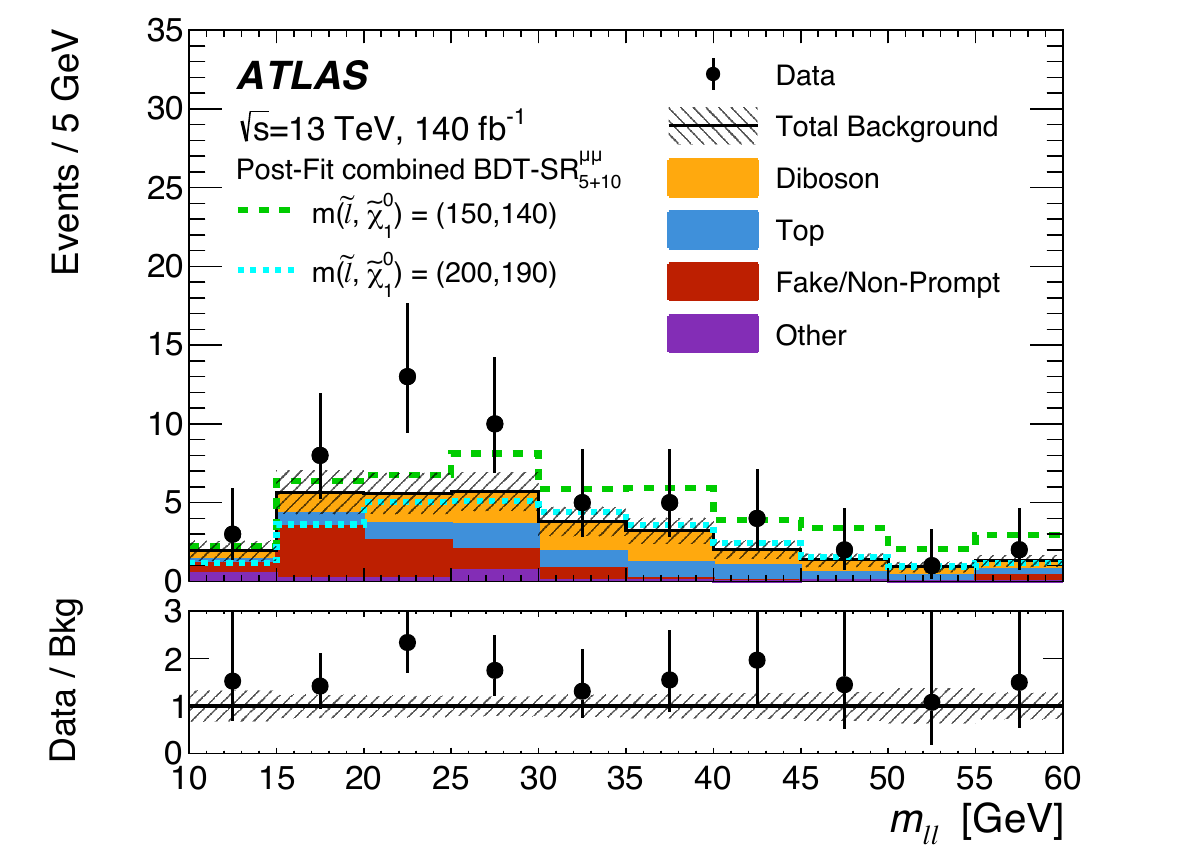}
}
\subfloat[]{
\includegraphics[width=0.49\textwidth]{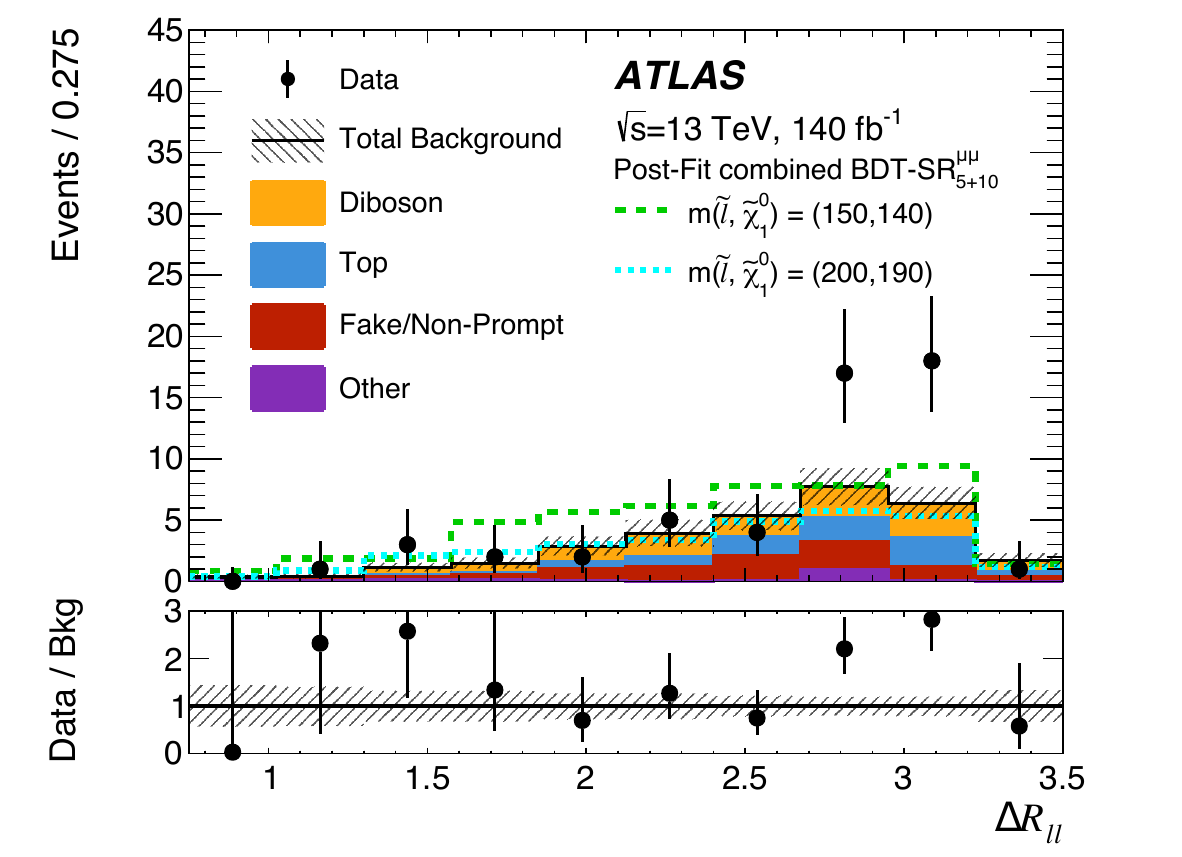}
}
\\
\subfloat[]{
\includegraphics[width=0.49\textwidth]{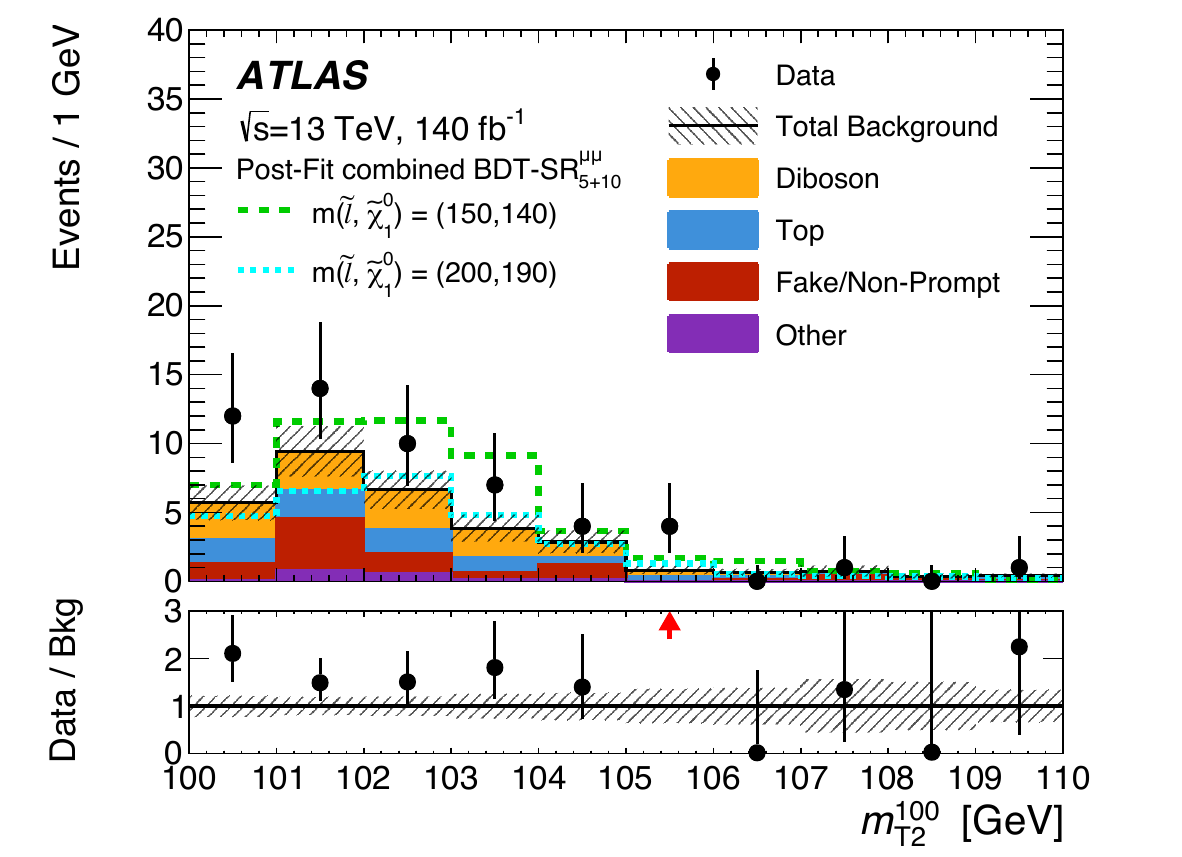}
}
\subfloat[]{
\includegraphics[width=0.49\textwidth]{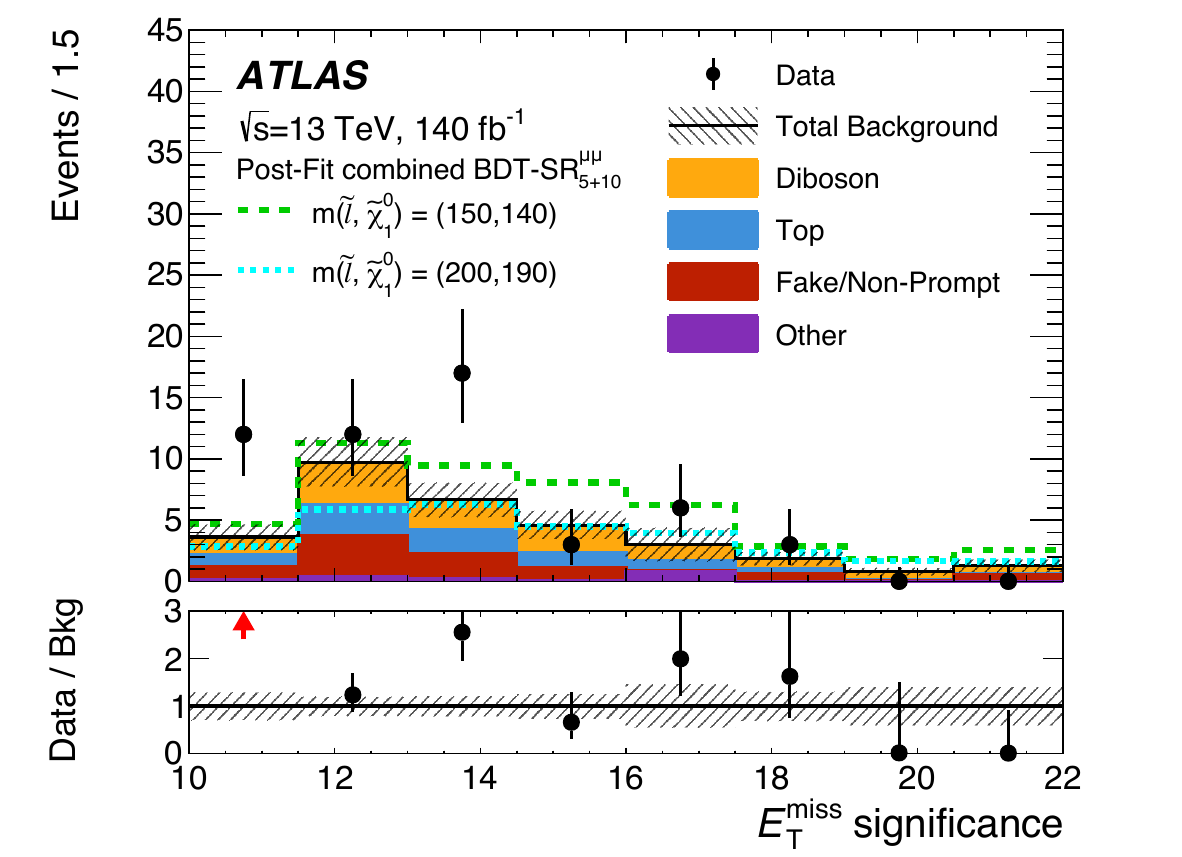}
}
\caption{Data (dots) and post-fit SM (histograms) distributions for (a) \mll, (b) \dRll, (c) $\mttwo^{100}$ and (d) the \metsig in the combined $\BDTSRmm_{5+10}$.
The uncertainty bands include statistical and systematic uncertainties.
The first and last bin include the under- and overflow (if applicable), respectively.
The lower panel shows the ratio of observed data to the total post-fit SM prediction.
The dashed lines show predicted yields for two benchmark signal models corresponding to $m(\slep, \chioz) = (150, 140)~\GeV$ and $(200, 190)~\GeV$, assuming $m(\slepL) = m(\slepR)$.
A red arrow in the lower panel indicates data points outside the vertical range shown.}
\label{fig:LowDMExcess}
\end{figure}

\subsection{Interpretations}

The observations in the SRs can be translated into constraints on the simplified models of slepton pair production.
These interpretations are performed for three different scenarios. The first scenario assumes, for the sake of simplicity, the four-fold mass degeneracy $m(\selectronL)=m(\selectronR)=m(\smuonL)=m(\smuonR)$ that also served as benchmark model for optimising the searches. The other two scenarios consider either selectron- or smuon-only production with $m(\selectron)\equiv m(\selectronL)=m(\selectronR)$ and $m(\smuon) \equiv m(\smuonL)=m(\smuonR)$, respectively. In the selectron-only scenario, the smuons are assumed to be decoupled with the rest of the SUSY sparticle spectrum and vise versa. Hence, only the $ee$ ($\mu\mu$) SRs are considered when constraining the selectron-only (smuon-only) scenario.

To determine the probability that the observed data are compatible with a particular signal model, the exclusion fit configuration is used in which the SRs are fitted simultaneously with the CRs.
The signal strength is a free parameter in these fits that coherently scales the nominal signal predictions across all regions.
The \cls prescription is then used to perform hypothesis tests and set exclusion limits at 95\% CL\@.
The exclusion limits are derived individually for the cut-and-count and each of the BDT-signal selections, as these are not statistically independent.
This renders six sets of exclusion limits in total, one cut-and-count-based and one for each of the BDT-based SRs.
To simplify the visualisation of the BDT-based results, these are combined into one set of exclusion constraints using the following prescription. For each signal model in the $\mslep$ -- $\dm$ mass plane, the BDT result yielding the lowest expected \cls value is used. This set of \cls values defines then the final BDT exclusion contour.

Considering first the signal models in which a four-fold slepton mass degeneracy is assumed,
Figure~\ref{fig:CombLimsSummary} shows the observed and expected cut-and-count and BDT-based exclusion limits.
The slepton-mass exclusion contours are shown in the range $2~\GeV \leq \dm  \leq 120~\GeV$ due to the presence of strong pre-existing ATLAS constraints at larger~\cite{SUSY-2013-11,SUSY-2018-32}, and smaller~\cite{SUSY-2018-16} mass splittings.
Scenarios with $\mslep < 95~\GeV$ are not considered, as the LEP experiments generally sets strong exclusion for these small slepton masses~\cite{LEP-slepton,ALEPH:2002nwp}.
The region excluded by the different interpretations is that bounded by the two exclusion contours, where two contours exist.
Both cut-and-count and BDT-based exclusion reaches are reduced at around $\dm=40~\GeV$ for this scenario. This originates from the excesses observed in \HighSRThreee in the BDT-based and \SRhighm in the cut-and-count-based selections, i.e. from electron and muon channels, respectively.
The increased sensitivity of the BDTs over the cut-and-count-based approach originates in particular from training multiple BDTs to optimise individually for a certain \dm, collectively providing high sensitivity across the full $\dM$ corridor.~\footnote{Training a BDT over the complete spectrum of mass splittings was found to yield sensitivity much closer to the cut-and-count-based analysis strategy.}
Besides effectively exploiting correlations between the input variables, also the looser lepton $\pt$ requirements applied at the preselection stage enable the BDT-based selections to retain sensitivity to scenarios with smaller mass splittings than those accessible with the cut-and-count approach.

The results can also be interpreted within models where $m(\selectronL) = m(\selectronR)$ and $m(\smuonL) = m(\smuonR)$, while the mass degeneracy between the slepton flavours is broken.
Specifically, the \selectronLR (\smuonLR) particles are assumed to be within the reach of the search, while the \smuonLR (\selectronLR) particles are decoupled at inaccessibly large masses.
The expected and observed exclusions for the BDT-based analysis within these models are shown in Figure~\ref{fig:BDTOnlyLimits}.
It can be observed that for both the selectron and smuon cases, the expected exclusion for the BDTs collectively spans the full range of $\dM$, across the challenging ``corridor'' region in which no previous experiment had sensitivity.
Again, the observed exclusions are weaker where excesses in the data were observed compared with the predictions of the background-only model, now with di-electron excesses affecting only the observed selectron exclusion, and correspondingly di-muon excesses that for smuons.

\begin{figure}[tbp]
\centering
\subfloat[]{
\includegraphics[width=0.65\textwidth]{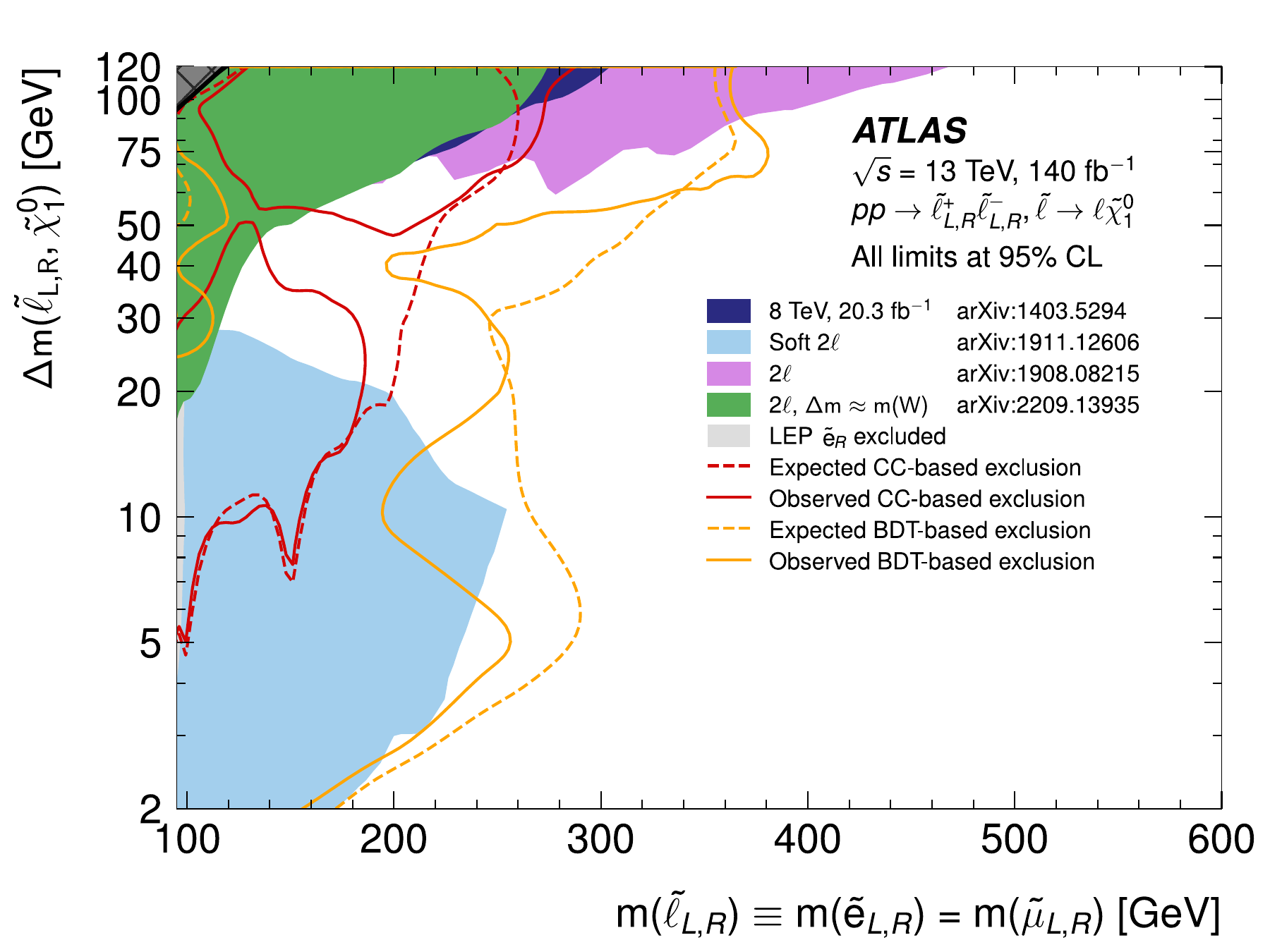}
}
\\
\subfloat[]{
\includegraphics[width=0.65\textwidth]{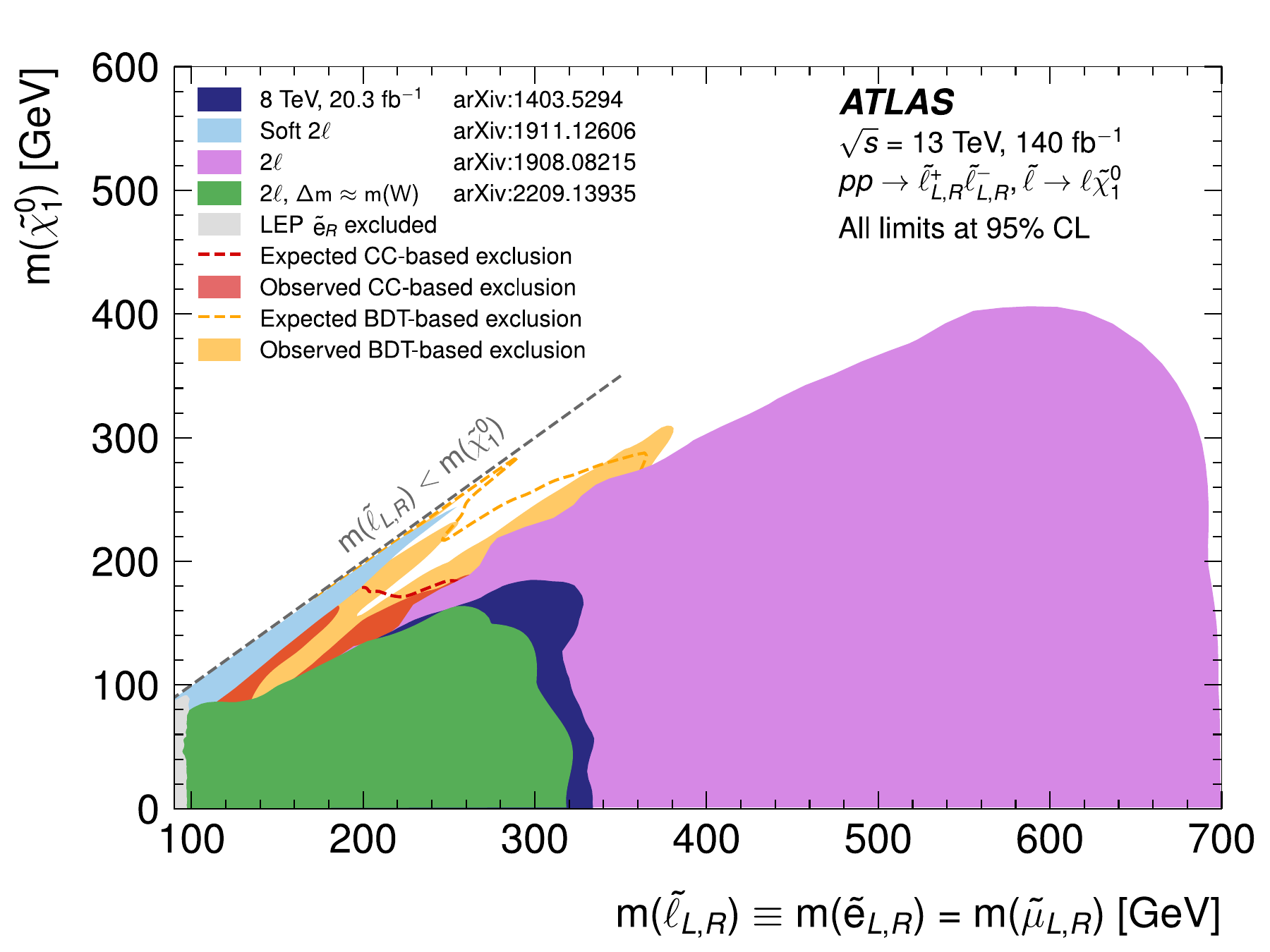}
}
\caption{Exclusion limits for simplified SUSY models of direct slepton production assuming four-fold mass degeneracy: $m(\slepton)\equiv m(\selectronL)=m(\selectronR)=m(\smuonL)=m(\smuonR)$.
The limits are shown in (a) the $m(\slep)$ -- $\Delta m(\slep, \tilde{\chi}_1^0)$ plane and (b) the $m(\slep)$ -- $m(\tilde{\chi}_1^0)$ plane.
In (a), a comparison is shown between the observed (orange solid lines) and expected (orange dashed lines) exclusion limits from the BDT approach, and the observed (solid red) and expected (dashed red) exclusion limits obtained from the cut-and-count-based approach.
In (b), a comparison is shown between the observed (orange fill) and expected (orange dashed lines) exclusion limits from the BDT approach, and the observed (red fill) and expected (dashed red) exclusion limits obtained from the cut-and-count-based approach.
All limits are computed at 95$\%$ CL, and the black delineated grey region indicates the kinematically forbidden region where $m(\tilde{\ell}_{L,R}) < m(\tilde{\chi}^0_1)$.
The observed limits obtained at LEP~\cite{LEP-slepton}, along with previous searches at the ATLAS experiment ~\cite{SUSY-2018-16,SUSY-2013-11,SUSY-2018-32,SUSY-2019-02} are shown.}
\label{fig:CombLimsSummary}
\end{figure}

\FloatBarrier

The data excess in \HighSRThreee results in a reduction in the observed exclusion limit for selectrons with $\dMselectron = $ 30--50~\GeV, preventing any exclusion of signal models within this region.
This results in a split of the exclusion contours into two ``islands'' of sensitivity for mass splittings below and above this region.
Stringent constraints are placed on selectrons in scenarios with \dMselectron between 2~\GeV and 20~\GeV, representing significant improvements over previous ATLAS results.
Notably, sensitivity is achieved at $\dMselectron = 20~\GeV$, extending the exclusion reach up to $\mselectron = 200~\GeV$, where the most stringent constraints were still from the LEP experiments~\cite{LEP-slepton}.
Enhanced exclusion limits are observed for mass splittings between 5~\GeV and 10~\GeV, up to a similar selectron mass.
Larger mass splittings with \dMselectron between 60~\GeV and 75~\GeV are constrained  up to $\mselectron = 300~\GeV$, again improving substantially over previous results.

For the smuon exclusion limits, a sizeable reduction in the observed excluded region compared with the expected one is seen for $\dMsmuon =$ 3--10~\GeV, driven by excesses in \VLowSROnem and \VLowSRTwom.
Except where excesses in the numbers of observed events reduce the ability to exclude models, significant improvements in the exclusion reach are observed for \dMsmuon between 20~\GeV and 100~\GeV, extending from $\smuonLR = 250~\GeV$ up to $\smuonLR = 350~\GeV$.

To identify which signal model would fit to the observed excesses best, the $p_{0}$-value of the background-only hypothesis can be evaluated relative to each signal model and translated into a discovery significance.
In contrast to the previously stated single-bin significances, all three bins of a BDT SR are included, which renders the hypothesis test sensitive to the signal shape.
The ambiguity over which set of BDT SRs is used for the evaluation is resolved by deriving the observed discovery significance with the set of BDT SRs that provides the best expected discovery significance for the signal model under consideration.
The discovery significances reported below are evaluated using 15,000 pseudo-experiments to render the distribution of the underlying test statistic.

For the two-fold degenerate selectron signal model the highest discovery significance is found to be $1.9\sigma$ and corresponds to the scenario with $\dMselectron = (375,335)~\GeV$ evaluated with $\HighSR^{ee}$.
The smuon model with the same masses, $\dMsmuon = (375,335)~\GeV$, has a discovery significance of $0.74\sigma$.
These discovery significances, evaluated relative to a particular signal model, are lower than the single-bin significances reported above
due to differences between the shapes of the signal predictions and the excesses observed in the SRs.
For the two-fold degenerate smuon signal model the highest discovery significance is provided by the $\dMsmuon = (150,140)~\GeV$ scenario using $\VLowSR^{\mu\mu}$ and amounts to $2.1\sigma$.~\footnote{When discounting signal models excluded by the LEP experiment.}
The selectron model with the same masses, $\dMselectron = (150,140)~\GeV$, yields a discovery significance of $0.18\sigma$.

\begin{figure}[tbp]
\centering
\subfloat[]{
\includegraphics[width=0.65\textwidth]{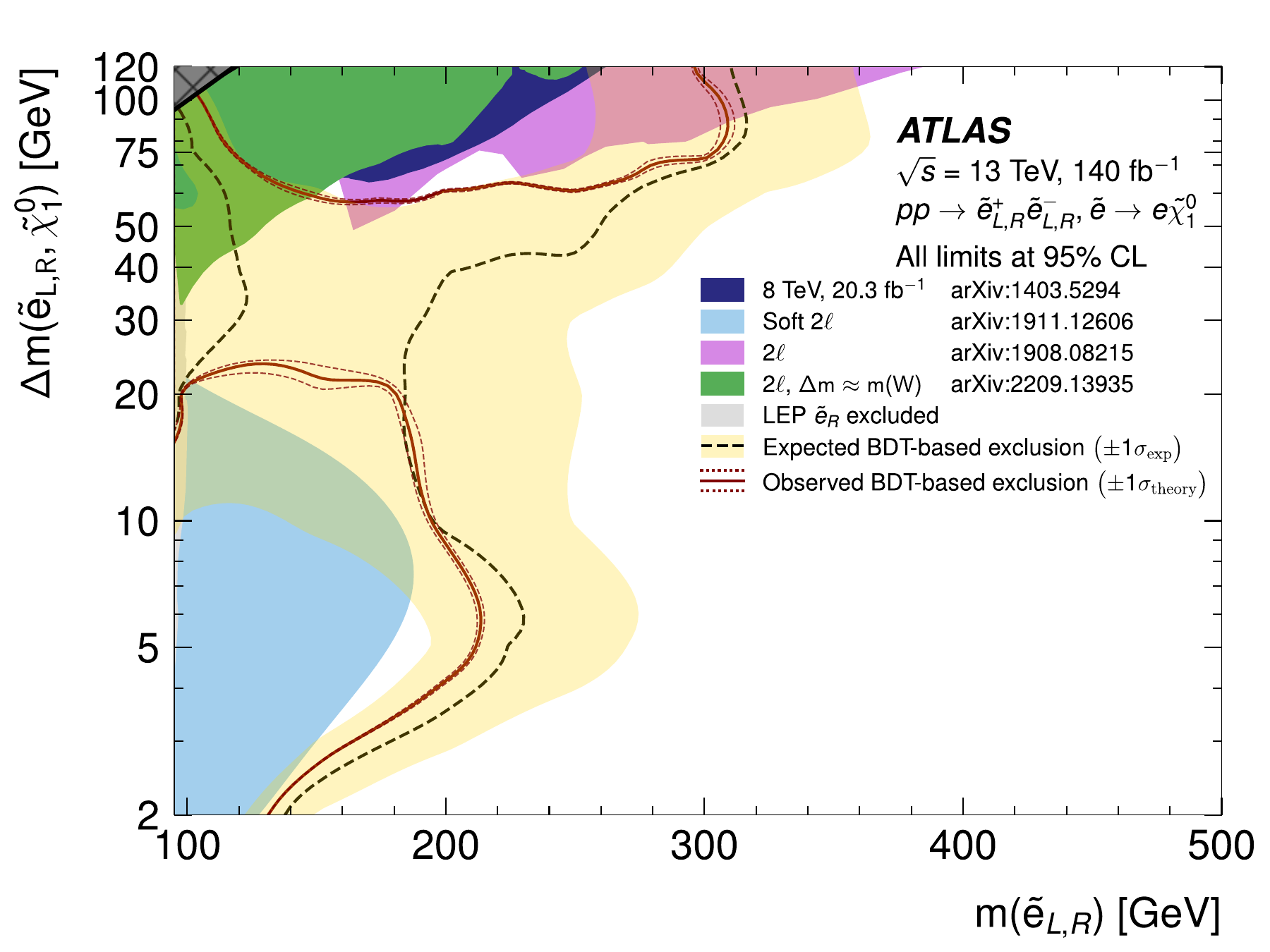}
}
\\
\subfloat[]{
\includegraphics[width=0.65\textwidth]{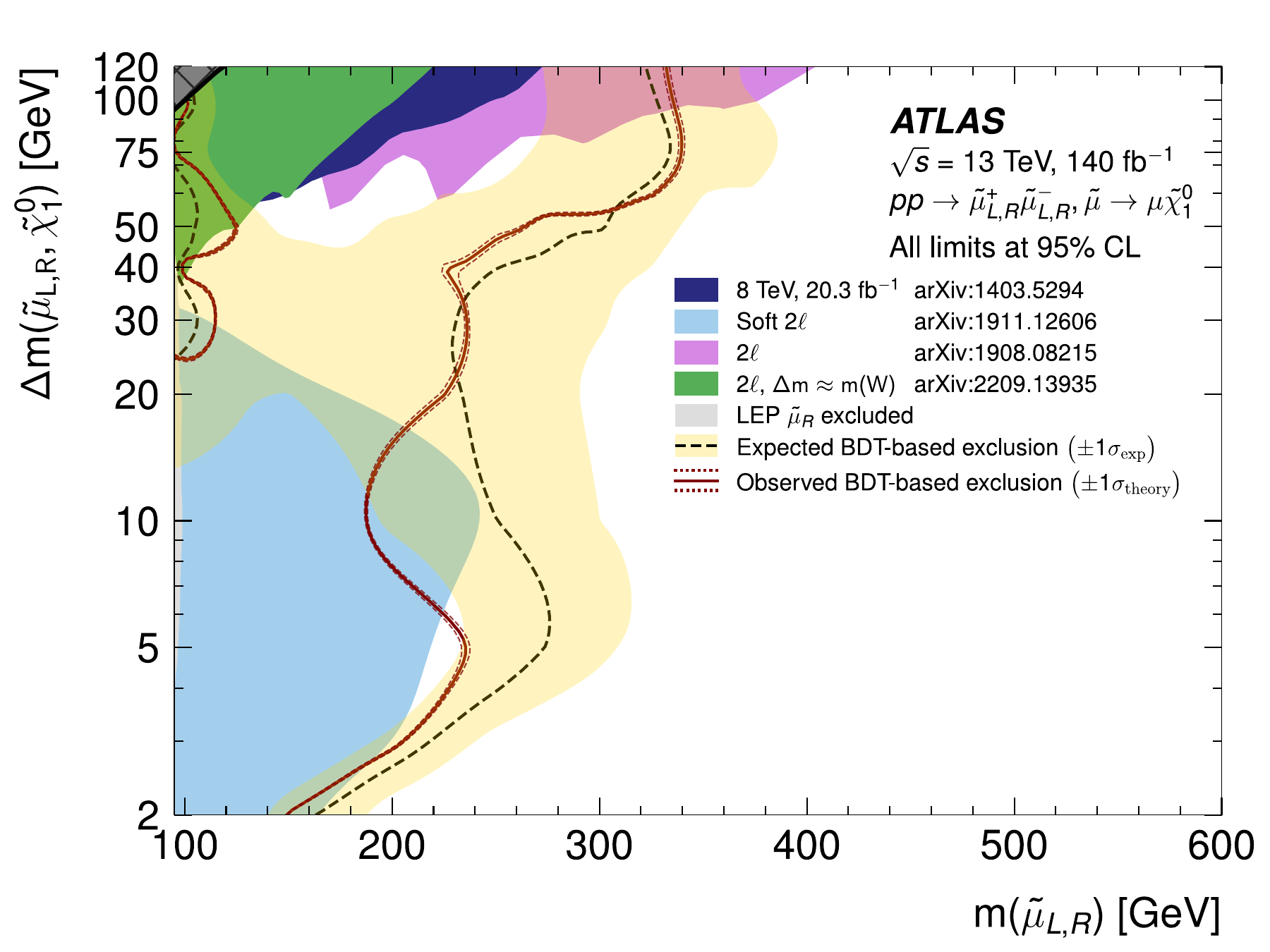}
}
\caption{Exclusion limits for simplified SUSY models assuming two-fold mass degeneracy. Separate limits are shown for (a) direct selectron production, where $m(\selectron) \equiv m(\selectronL) = m(\selectronR)$, and (b) direct smuon production, where $m(\smuon) \equiv m(\smuonL) = m(\smuonR)$.
All limits are computed at 95$\%$ CL and the observed (red solid lines) and expected (black dashed lines) exclusion limits from the BDT regions are shown in the $m(\tilde{e})$ -- $\Delta m(\tilde{e}, \tilde{\chi}_1^0)$ and  $m(\tilde{\mu})$ -- $\Delta m(\tilde{\mu}, \tilde{\chi}_1^0)$ planes, respectively.
In (a), the observed exclusion should be interpreted as the region enclosed by each separate observed curve and the axis.
For the expected exclusion in (a) and the expected and observed exclusion in (b), the excluded region should be interpreted as the region enclosed by the two observed curves.
The black delineated grey region indicates the kinematically forbidden region where $m(\selectronLR / \smuonLR) < m(\tilde{\chi}^0_1)$.
The yellow shaded band around the expected limits corresponds to the $\pm 1\,\sigma$ variations of the expected limit, accounting for all uncertainties aside from the theoretical uncertainties in the signal cross-section.
The dotted lines around the observed exclusion limits indicate the variation as the nominal signal cross-section is scaled up and down by the theoretical uncertainty.
The observed limits obtained at LEP ~\cite{LEP-slepton,ALEPH:2002nwp}, along with previous searches at the ATLAS experiment ~\cite{SUSY-2018-16,SUSY-2013-11,SUSY-2018-32,SUSY-2019-02} are shown.}
\label{fig:BDTOnlyLimits}
\end{figure}

\FloatBarrier

% End of text imported from the .//sleptons/results.tex input file

%
%
%
\FloatBarrier
\section{Conclusion}
%

% The next lines are included from the .//sleptons/conclusion.tex input file
%
%
Searches are performed for supersymmetric leptons, $\slepton$, of the first two generations. The sleptons, selectrons or smuons, are assumed to be pair-produced and to decay into a pair of charged leptons with opposite charge in the same family and two unobserved stable neutralinos, $\chioz$. A dedicated effort was made to be sensitive to models with small mass splitting between slepton and neutralino similar or less the mass of the $W$ boson, $\dm\lesssim m(W)$, where no searches so far have had sufficient sensitivity.
The searches use the complete LHC Run 2 data sample from the ATLAS experiment, equivalent to $140\,\ifb$ of integrated luminosity.
A \met{} trigger is employed, and a high \pT{} jet required, consistent with the emission of initial-state radiation, to evade constraints on lepton \pT{} that would arise from using lepton triggers.

Two complementary approaches are employed, a traditional cut-and-count search and an analysis based on boosted decision trees.
Each search places requirements on kinematic variables designed to be sensitive to both the masses and the spin structures of the production and decay processes, helping distinguish the slepton pair signal from the leptonic $W^+W^-$ background.
Good agreement is found between the predictions and observed yields in a wide range of control and validation regions.
The most sensitive results are found using a set of five boosted decision trees, each trained for a different region of $\dm$ in the range of 5~\GeV to 75\,\GeV, with expected sensitivity to $\smuon$ or $\selectron$ masses ranging from 200~\GeV to 350\,\GeV for the benchmark SUSY models with $\dm$ larger than $2\,\GeV$, providing expected sensitivity across the full $\dM$ corridor, even for scenarios in which four-fold mass degeneracy of the sleptons is not assumed.

The results are generally consistent with the Standard Model expectations for both the cut-and-count-based and the BDT-based searches.
The most significant excesses are found in the BDT-based searches with a local significance of $2.0\,\sigma$ in the search for selectrons and $2.4\,\sigma$ in the search for smuons.
Limits are set at the 95\% CL on $\smuon$ masses up to 350\,\GeV for particular values of $\dm$, assuming $m(\smuonL)=m(\smuonR)$ and similarly on $\selectron$ masses up to 300\,\GeV assuming $m(\selectronL)=m(\selectronR)$.
Thus, this analysis significantly strengthens the constraints on these benchmark models compared to previous searches and largely addresses a long-standing sensitivity gap in the \dM corridor at moderately compressed mass splittings.
Notably, for scenarios with $m(\smuonL)=m(\smuonR)$ and mass splittings \dMsmuon between 30--40~\GeV, where the strongest limits were previously set by the LEP experiments, the exclusion is extended from around $\msmuon > 100~\GeV$ to approximately $\msmuon > 250~\GeV$.
The regions of slepton and neutralino mass that are excluded are smaller than expected
due to the observed excesses,
particularly for smuons with $\dm\approx\,10\,\GeV$ and selectrons with $\dm\approx\,40\,\GeV$.
The maximum discovery significance found is $1.9\,\sigma$ for the selectron signal model with $m(\selectron, \chioz)=(375,335)~\GeV$, and $2.1\,\sigma$ for the smuon signal model with $m(\smuon,\chioz)=(150,140)~\GeV$.

% End of text imported from the .//sleptons/conclusion.tex input file

%
%
%
%
%
\section*{Acknowledgements}
%

%

% The next lines are included from the .//acknowledgements/Acknowledgements.tex input file
%
%

%
%

We thank CERN for the very successful operation of the LHC and its injectors, as well as the support staff at
CERN and at our institutions worldwide without whom ATLAS could not be operated efficiently.

The crucial computing support from all WLCG partners is acknowledged gratefully, in particular from CERN, the ATLAS Tier-1 facilities at TRIUMF/SFU (Canada), NDGF (Denmark, Norway, Sweden), CC-IN2P3 (France), KIT/GridKA (Germany), INFN-CNAF (Italy), NL-T1 (Netherlands), PIC (Spain), RAL (UK) and BNL (USA), the Tier-2 facilities worldwide and large non-WLCG resource providers. Major contributors of computing resources are listed in Ref.~\cite{ATL-SOFT-PUB-2025-001}.

We gratefully acknowledge the support of ANPCyT, Argentina; YerPhI, Armenia; ARC, Australia; BMWFW and FWF, Austria; ANAS, Azerbaijan; CNPq and FAPESP, Brazil; NSERC, NRC and CFI, Canada; CERN; ANID, Chile; CAS, MOST and NSFC, China; Minciencias, Colombia; MEYS CR, Czech Republic; DNRF and DNSRC, Denmark; IN2P3-CNRS and CEA-DRF/IRFU, France; SRNSFG, Georgia; BMBF, HGF and MPG, Germany; GSRI, Greece; RGC and Hong Kong SAR, China; ICHEP and Academy of Sciences and Humanities, Israel; INFN, Italy; MEXT and JSPS, Japan; CNRST, Morocco; NWO, Netherlands; RCN, Norway; MNiSW, Poland; FCT, Portugal; MNE/IFA, Romania; MSTDI, Serbia; MSSR, Slovakia; ARIS and MVZI, Slovenia; DSI/NRF, South Africa; MICIU/AEI, Spain; SRC and Wallenberg Foundation, Sweden; SERI, SNSF and Cantons of Bern and Geneva, Switzerland; NSTC, Taipei; TENMAK, T\"urkiye; STFC/UKRI, United Kingdom; DOE and NSF, United States of America.

Individual groups and members have received support from BCKDF, CANARIE, CRC and DRAC, Canada; CERN-CZ, FORTE and PRIMUS, Czech Republic; COST, ERC, ERDF, Horizon 2020, ICSC-NextGenerationEU and Marie Sk{\l}odowska-Curie Actions, European Union; Investissements d'Avenir Labex, Investissements d'Avenir Idex and ANR, France; DFG and AvH Foundation, Germany; Herakleitos, Thales and Aristeia programmes co-financed by EU-ESF and the Greek NSRF, Greece; BSF-NSF and MINERVA, Israel; NCN and NAWA, Poland; La Caixa Banking Foundation, CERCA Programme Generalitat de Catalunya and PROMETEO and GenT Programmes Generalitat Valenciana, Spain; G\"{o}ran Gustafssons Stiftelse, Sweden; The Royal Society and Leverhulme Trust, United Kingdom.

In addition, individual members wish to acknowledge support from Armenia: Yerevan Physics Institute (FAPERJ); CERN: European Organization for Nuclear Research (CERN DOCT); Chile: Agencia Nacional de Investigaci\'on y Desarrollo (FONDECYT 1230812, FONDECYT 1240864); China: Chinese Ministry of Science and Technology (MOST-2023YFA1605700, MOST-2023YFA1609300), National Natural Science Foundation of China (NSFC - 12175119, NSFC 12275265); Czech Republic: Czech Science Foundation (GACR - 24-11373S), Ministry of Education Youth and Sports (ERC-CZ-LL2327, FORTE CZ.02.01.01/00/22\_008/0004632), PRIMUS Research Programme (PRIMUS/21/SCI/017); EU: H2020 European Research Council (ERC - 101002463); European Union: European Research Council (BARD No. 101116429, ERC - 948254, ERC 101089007), European Regional Development Fund (SMASH COFUND 101081355, SLO ERDF), Horizon 2020 Framework Programme (MUCCA - CHIST-ERA-19-XAI-00), European Union, Future Artificial Intelligence Research (FAIR-NextGenerationEU PE00000013), Horizon 2020 (EuroHPC - EHPC-DEV-2024D11-051), Italian Center for High Performance Computing, Big Data and Quantum Computing (ICSC, NextGenerationEU); France: Agence Nationale de la Recherche (ANR-21-CE31-0013, ANR-21-CE31-0022, ANR-22-EDIR-0002); Germany: Baden-Württemberg Stiftung (BW Stiftung-Postdoc Eliteprogramme), Deutsche Forschungsgemeinschaft (DFG - 469666862, DFG - CR 312/5-2); China: Research Grants Council (GRF); Italy: Istituto Nazionale di Fisica Nucleare (ICSC, NextGenerationEU), Ministero dell'Università e della Ricerca (NextGenEU I53D23000820006 M4C2.1.1); Japan: Japan Society for the Promotion of Science (JSPS KAKENHI JP22H01227, JSPS KAKENHI JP22H04944, JSPS KAKENHI JP22KK0227, JSPS KAKENHI JP23KK0245, JSPS KAKENHI JP24K23939); Norway: Research Council of Norway (RCN-314472); Poland: Ministry of Science and Higher Education (IDUB AGH, POB8, D4 no 9722), Polish National Science Centre (NCN 2021/42/E/ST2/00350, NCN OPUS 2023/51/B/ST2/02507, NCN OPUS nr 2022/47/B/ST2/03059, NCN UMO-2019/34/E/ST2/00393, UMO-2020/37/B/ST2/01043, UMO-2022/47/O/ST2/00148, UMO-2023/49/B/ST2/04085, UMO-2023/51/B/ST2/00920, UMO-2024/53/N/ST2/00869); Portugal: Foundation for Science and Technology (FCT); Spain: Generalitat Valenciana (Artemisa, FEDER, IDIFEDER/2018/048), Ministry of Science and Innovation (MCIN \& NextGenEU PCI2022-135018-2, MICIN \& FEDER PID2021-125273NB, RYC2019-028510-I, RYC2020-030254-I, RYC2021-031273-I, RYC2022-038164-I); Sweden: Carl Trygger Foundation (Carl Trygger Foundation CTS 22:2312), Swedish Research Council (Swedish Research Council 2023-04654, VR 2021-03651, VR 2022-03845, VR 2022-04683, VR 2023-03403, VR 2024-05451), Knut and Alice Wallenberg Foundation (KAW 2018.0458, KAW 2022.0358, KAW 2023.0366); Switzerland: Swiss National Science Foundation (SNSF - PCEFP2\_194658); United Kingdom: Leverhulme Trust (Leverhulme Trust RPG-2020-004), Royal Society (NIF-R1-231091); United States of America: U.S. Department of Energy (ECA DE-AC02-76SF00515), Neubauer Family Foundation.

%
%

% End of text imported from the .//acknowledgements/Acknowledgements.tex input file

%
\clearpage
\appendix
\part*{Appendix}
\addcontentsline{toc}{part}{Appendix}

% The next lines are included from the .//sleptons/appendix.tex input file

\FloatBarrier
\section{BDT Training}
\label{sec:BDTTraining}

Signal samples with the same mass-splittings share similar kinematic features, and hence are grouped together during training to enhance the number of signal events in the training data.
The final grouping of the individual signal samples is shown for the full grid of signal samples in Figure~\ref{fig:SignalGrouping}.
Every signal in each grouping is given a proportional weighting within the training of each BDT\@.
Some signals are not trained on as they were not the region which the analysis set out to target.

\begin{figure}[bp]
\centering
\includegraphics[width=0.9\textwidth]{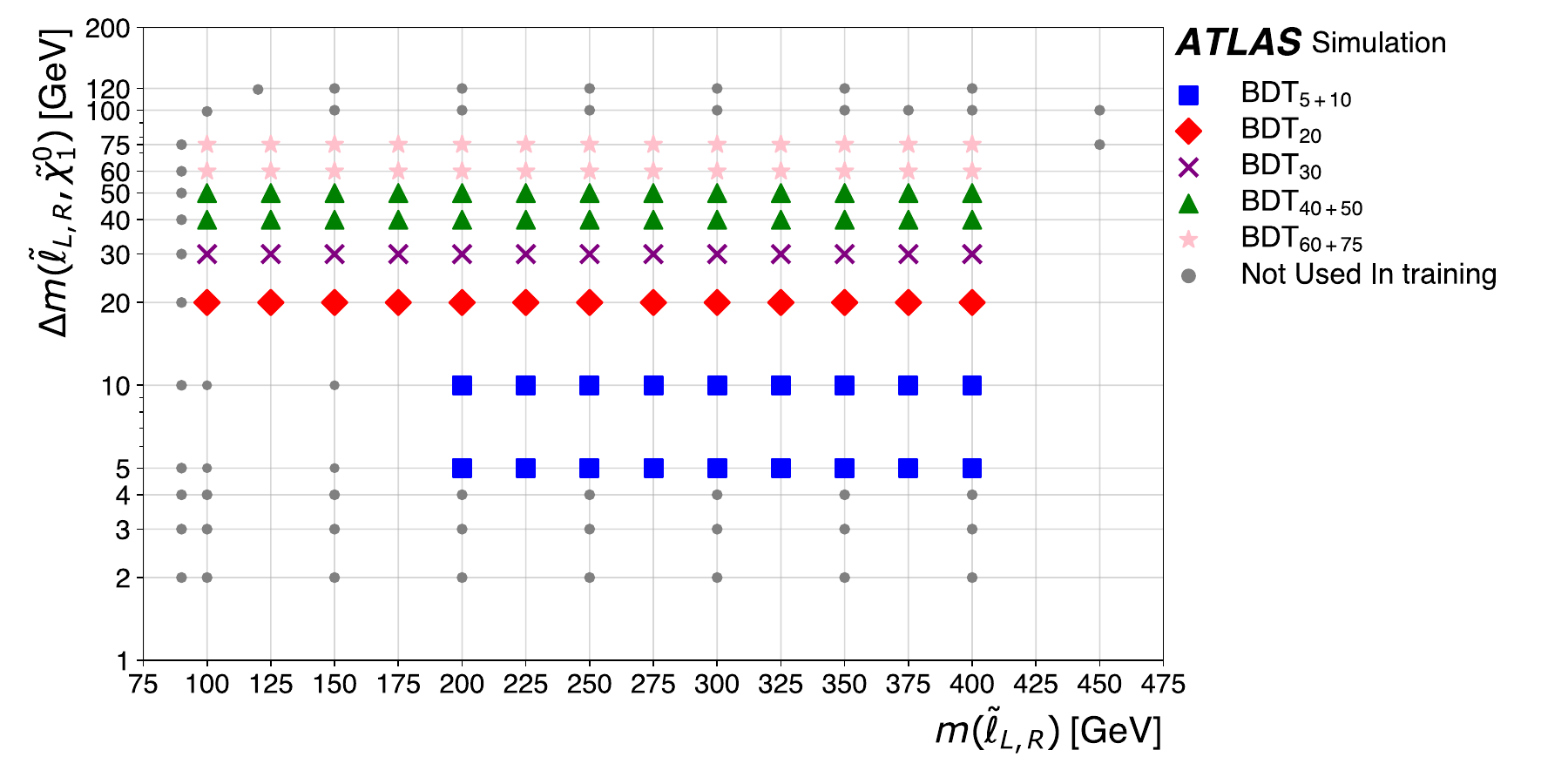}
\caption{Overview in the $\mslepLR$ -- $\dMslepLR$ plane of how the simulated benchmark signal models are grouped for the BDT training. The samples entering into each BDT are colour and symbol coded.}
\label{fig:SignalGrouping}
\end{figure}

Hyperparameter optimisation was carried out independently for all five \dM BDTs.
The following hyperparameters were varied:
\begin{itemize}
\item The maximum number of trees in the BDT -- this does not get reached by design due to early stopping.
\item The learning rate -- a scale factor determining how much the loss is modified after each iteration
\item The minimum number of events per leaf (minimum child weight) -- number of events before a node splits
\item The Gamma parameter -- implements the pruning (dropping) of leaves to control tree complexity and prevent overtraining
\item The Alpha Regularisation parameter -- a penalty term to the gradient of the loss function prevent overtraining.
\item The Lambda Regularisation -- a penalty term to the Hessian of the loss function prevent overtraining.
\end{itemize}
The same hyperparameter configurations are used for all five different k-fold trainings, with the final number of trees varying by 1--2\% between the different k-fold trainings.
The final hyperparameter configurations for each of the five BDTs are detailed in Table \ref{tab:BDTHyper}.

% The next lines are included from the .//tables/BDTHyperparameters.tex input file
\begin{table}[tbp]
\centering
\renewcommand{\arraystretch}{1.1}
\caption{Hyperparameter choices for each BDT\@. The final number of trees varies across the different folds of the same BDT and so the average is taken.}
\begin{tabular}[h]{cccccc}
\toprule
\toprule
Hyperparameter & $\mathrm{BDT_{5+10}}$ & $\mathrm{BDT_{20}}$ & $\mathrm{BDT_{30}}$ & $\mathrm{BDT_{40+50}}$ & $\mathrm{BDT_{60+75}}$\\
\midrule
(Average) final number of trees & 372 & 489& 668 & 789 & 924\\
Learning rate &0.025&0.025&0.04&0.025&0.025 \\
Minimum number of events per leaf &5&2&5&2&15 \\
Alpha &6&1&1&1&1 \\
Gamma &1.1&0.4&0.3&0.25&0.25 \\
Lambda &12&5&6&20&15 \\
\bottomrule
\bottomrule
\end{tabular}
\label{tab:BDTHyper}
\end{table}

% End of text imported from the .//tables/BDTHyperparameters.tex input file

The three most significant variables influencing the training of each BDT, as evaluated using the SHAP package~\cite{SHAP}, are summarised in Table \ref{tab:VarImp}.
The kinematics of the signal depend strongly on the \dm of the signal, leading to substantial variation in the most important variables across the BDTs.
For the smallest \dm, the most important variable is \mttwo, where low values are used to exploit the kinematic endpoint of the signal models.
Specifically, the importance of the large invisible mass variants of \mttwo are indicative of the fact that this BDT was exclusively trained using signals with $m_{\chi}>190~\GeV$.
Additionally, low \ptll is a key variable, used to effectively discriminating between signal leptons and FNP leptons.
For the largest \dm BDT, the most important variables are \mtleptwo, where high values play a crucial role.
Large $\metsig$ is used to select signal events, which are characterised by large values of real \MET.

% The next lines are included from the .//tables/BDTVariableImportance.tex input file
\begin{table}[tbp]
\centering
\renewcommand{\arraystretch}{1.1}
\caption[]{The three most important variables during training for each BDT using the metrics from the SHAP package.}
\begin{tabular}[h]{lccc}
\toprule
\toprule
BDT & First ranked & Second ranked & Third ranked\\
\hline
$\mathrm{BDT}_{5+10}$ & $\mttwo^{300}$ & $\mttwo^{200}$ & $\ptll$\\
$\mathrm{BDT}_{20}$  & $\mttwo^{200}$ & $\mttwo^{150}$ &\metsig\\
$\mathrm{BDT}_{30}$  & \metsig & \mtleptwo & $\mttwo^{200}$\\
$\mathrm{BDT}_{40+50}$ & \mtleptwo &  \metsig & \mtlepone \\
$\mathrm{BDT}_{60+75}$ & \mtleptwo &  \metsig & \mll \\
\bottomrule
\bottomrule
\end{tabular}
\label{tab:VarImp}
\end{table}

% End of text imported from the .//tables/BDTVariableImportance.tex input file

In the training, \mll approximates $2\times \dm$ for the signal models, and at the largest \dm, large values of \mll are utilised to select signal events.
For the intermediate \dm BDTs, the most important variables consist of a combination of those relevant to both the lowest and highest \dm BDTs.
As the \dm of the signal models increases, the kinematics evolve from low-\pt leptons associated with high \MET, to leptons with large \mtleptwo and \mll.
Consequently, the most important variables for each BDT reflect this behaviour.

% End of text imported from the .//sleptons/appendix.tex input file

%
%
%
\printbibliography
\clearpage
% ATLAS Collaboration author list
% Reference date of HMBS-2024-64 is 2025-01-24
% Author list last updated on date 20-JUN-25
% Data extracted on 20-Jun-2025 for paper reference HMBS-2024-64
% at 2:32pm
 
\begin{flushleft}
\hypersetup{urlcolor=black}
{\Large The ATLAS Collaboration}

\bigskip

\AtlasOrcid[0000-0002-6665-4934]{G.~Aad}$^\textrm{\scriptsize 104}$,
\AtlasOrcid[0000-0001-7616-1554]{E.~Aakvaag}$^\textrm{\scriptsize 17}$,
\AtlasOrcid[0000-0002-5888-2734]{B.~Abbott}$^\textrm{\scriptsize 123}$,
\AtlasOrcid[0000-0002-0287-5869]{S.~Abdelhameed}$^\textrm{\scriptsize 119a}$,
\AtlasOrcid[0000-0002-1002-1652]{K.~Abeling}$^\textrm{\scriptsize 55}$,
\AtlasOrcid[0000-0001-5763-2760]{N.J.~Abicht}$^\textrm{\scriptsize 49}$,
\AtlasOrcid[0000-0002-8496-9294]{S.H.~Abidi}$^\textrm{\scriptsize 30}$,
\AtlasOrcid[0009-0003-6578-220X]{M.~Aboelela}$^\textrm{\scriptsize 45}$,
\AtlasOrcid[0000-0002-9987-2292]{A.~Aboulhorma}$^\textrm{\scriptsize 36e}$,
\AtlasOrcid[0000-0001-5329-6640]{H.~Abramowicz}$^\textrm{\scriptsize 157}$
% ...
\bigskip
\\

% ...
$^{17}$Department for Physics and Technology, University of Bergen, Bergen; Norway.\\
% ...
$^{30}$Physics Department, Brookhaven National Laboratory, Upton NY; United States of America.\\
% ...
$^{36}$$^{(a)}$Facult\'e des Sciences Ain Chock, Universit\'e Hassan II de Casablanca;$^{(b)}$Facult\'{e} des Sciences, Universit\'{e} Ibn-Tofail, K\'{e}nitra;$^{(c)}$Facult\'e des Sciences Semlalia, Universit\'e Cadi Ayyad, LPHEA-Marrakech;$^{(d)}$LPMR, Facult\'e des Sciences, Universit\'e Mohamed Premier, Oujda;$^{(e)}$Facult\'e des sciences, Universit\'e Mohammed V, Rabat;$^{(f)}$Institute of Applied Physics, Mohammed VI Polytechnic University, Ben Guerir; Morocco.\\
% ...
$^{45}$Physics Department, Southern Methodist University, Dallas TX; United States of America.\\
% ...
$^{49}$Fakult\"{a}t Physik , Technische Universit{\"a}t Dortmund, Dortmund; Germany.\\
% ...
$^{55}$II. Physikalisches Institut, Georg-August-Universit\"{a}t G\"ottingen, G\"ottingen; Germany.\\
% ...
$^{104}$CPPM, Aix-Marseille Universit\'e, CNRS/IN2P3, Marseille; France.\\
% ...
$^{119}$$^{(a)}$New York University Abu Dhabi, Abu Dhabi;$^{(b)}$United Arab Emirates University, Al Ain; United Arab Emirates.\\
% ...
$^{123}$Homer L. Dodge Department of Physics and Astronomy, University of Oklahoma, Norman OK; United States of America.\\
% ...
$^{157}$Raymond and Beverly Sackler School of Physics and Astronomy, Tel Aviv University, Tel Aviv; Israel.\\
% ...
$^{a}$ Also at Affiliated with an institute formerly covered by a cooperation agreement with CERN.\\
$^{b}$ Also at An-Najah National University, Nablus; Palestine.\\
$^{c}$ Also at Borough of Manhattan Community College, City University of New York, New York NY; United States of America.\\
$^{d}$ Also at Center for Interdisciplinary Research and Innovation (CIRI-AUTH), Thessaloniki; Greece.\\
$^{e}$ Also at Centre of Physics of the Universities of Minho and Porto (CF-UM-UP); Portugal.\\
$^{f}$ Also at CERN, Geneva; Switzerland.\\
% ...

\end{flushleft}

% Created with Glance <Atlas.Glance@cern.ch>

%

%
%

%

%
%
%
%

\end{document}